\documentclass[conference]{IEEEtran}
\usepackage[hyphens]{url}
\usepackage{hyperref}
\usepackage[hyphenbreaks]{breakurl}
\usepackage{cite}

\ifCLASSINFOpdf
\else
\fi

\usepackage{hyperref}
\usepackage{tikz}
\usetikzlibrary{patterns}
\usepackage{amsmath}
\usepackage{xspace}
\usepackage{cleveref}
\usepackage{adjustbox}
\usepackage{booktabs}
\usepackage{caption}
\usepackage{subcaption}
\usepackage{siunitx}[binary-units]
\DeclareSIUnit\byte{B}
\usepackage{pgfplots}
\usepackage{filecontents}
\usepackage{graphics}
\usepackage{colortbl}
\usepackage{listings}
\usepackage{pifont}
\usepackage{paralist} %
\usepackage{csquotes} %
\usepackage{fontawesome}

\usepackage{ltxcmds}
\makeatletter
\newcommand*{\getdata}[2]{%
  \expandafter\ltx@CarNumth\expandafter{%
    \the\numexpr(#1)\expandafter
  }#2\@nil
}
\makeatother

\newcommand{\ie}{\textit{i.e.},\ }
\newcommand{\eg}{e.g.,\ }
\newcommand{\Eg}{E.g.,\ }
\newcommand{\cf}{cf.\ }

\definecolor{fancyred}{HTML}{f70146}
\definecolor{fancypurple}{HTML}{6c2f91}
\definecolor{fancymid}{HTML}{5191c1}
\definecolor{fancygray}{HTML}{a5a5a5}
\definecolor{fancyblue}{HTML}{285f82}
\definecolor{fancygreen}{HTML}{78b473}
\definecolor{fancyyellow}{HTML}{e59352}
\definecolor{fancycyan}{HTML}{77babf}
\definecolor{fancyhead}{HTML}{245b78}
\definecolor{fancybody}{HTML}{e2e9ed}
\colorlet{cA2}{fancygray}
\colorlet{cA}{fancygray!50}
\colorlet{cB}{fancyblue}
\colorlet{cC}{fancygreen}
\colorlet{cD}{fancyyellow}
\colorlet{cE}{fancyred}
\colorlet{cF}{fancypurple}
\colorlet{cG}{fancycyan}
\colorlet{cH}{black}

\reversemarginpar

\renewcommand{\paragraph}[1]{\noindent \textbf{#1.}\ }

\newcommand{\tmebox}{TME-Box\xspace}

\newcommand{\keyid}{keyID\xspace}
\newcommand{\keyids}{keyIDs\xspace}
\newcommand{\tmemk}{TME-MK\xspace}

\newcommand{\SPECgs}{5.2\,\%\xspace}
\newcommand{\SPECgscfi}{9.7\,\%\xspace}
\newcommand{\SPECr}{13.4\,\%\xspace}
\newcommand{\SPECrcfi}{17.7\,\%\xspace}

\newcommand{\NGINXgsmax}{8.3\,\%\xspace}
\newcommand{\NGINXgsmin}{1.8\,\%\xspace}
\newcommand{\NGINXrmax}{8.7\,\%\xspace}
\newcommand{\NGINXrmin}{1.9\,\%\xspace}

\begin{document}
\title{TME-Box: Scalable In-Process Isolation through Intel TME-MK Memory Encryption}

\author{\IEEEauthorblockN{Martin Unterguggenberger\IEEEauthorrefmark{1},
Lukas Lamster\IEEEauthorrefmark{1},
David Schrammel\IEEEauthorrefmark{1},
Martin Schwarzl\IEEEauthorrefmark{2},
Stefan Mangard\IEEEauthorrefmark{1}}
\IEEEauthorblockA{\IEEEauthorrefmark{1}Graz University of Technology: \{firstname.lastname\}@iaik.tugraz.at}
\IEEEauthorblockA{\IEEEauthorrefmark{2}Cloudflare, Inc.: mschwarzl@cloudflare.com}}

\IEEEoverridecommandlockouts
\makeatletter\def\@IEEEpubidpullup{6.5\baselineskip}\makeatother
\IEEEpubid{\parbox{\columnwidth}{
		\phantom{Network and Distributed System Security (NDSS) Symposium 2025}\\
		\phantom{23–28 February 2025, San Diego, CA, USA}\\
		\phantom{ISBN 979-8-9894372-8-3}\\
		\phantom{https://dx.doi.org/10.14722/ndss.2025.24277}\\
		\phantom{www.ndss-symposium.org}
}
\hspace{\columnsep}\makebox[\columnwidth]{}}

\maketitle

\begin{abstract}

Efficient cloud computing relies on in-process isolation to optimize performance by running workloads within a single process.
Without heavy-weight process isolation, memory safety errors pose a significant security threat by allowing an adversary to extract or corrupt the private data of other co-located tenants.
Existing in-process isolation mechanisms are not suitable for modern cloud requirements, \eg MPK's 16 protection domains are insufficient to isolate thousands of cloud workers per process.
Consequently, cloud service providers have a strong need for lightweight in-process isolation on commodity x86 machines. %

This paper presents \tmebox, a novel isolation technique that enables fine-grained and scalable sandboxing on commodity x86 CPUs. %
By repurposing Intel \tmemk, which is intended for the encryption of virtual machines, \tmebox offers lightweight and efficient in-process isolation. %
\tmebox enforces that sandboxes use their designated encryption keys for memory interactions through compiler instrumentation.
This cryptographic isolation enables fine-grained access control, from single cache lines to full pages, and supports flexible data relocation.
In addition, the design of \tmebox allows the efficient isolation of up to 32K concurrent sandboxes.
We present a performance-optimized \tmebox prototype, utilizing x86 segment-based addressing, that showcases geomean performance overheads of \SPECgs for data isolation and \SPECgscfi for code and data isolation, evaluated with the SPEC CPU2017 benchmark suite.

\end{abstract}

\IEEEpeerreviewmaketitle

\section{Introduction}\label{sec:tmebox:introduction}

Cloud computing must be highly optimized for performance and efficiency, processing requests with the lowest possible latencies.
Various techniques and optimizations are implemented at the software and infrastructural level of the cloud architecture.
Common optimizations are the exclusion of protection mechanisms, like process isolation, in favor of low start-up times and fast execution for serverless applications in a multi-tenant environment, \eg function-as-a-service~(FaaS) applications~\cite{fastly,deno,CFWorkers}. %
However, omitting heavy-weight protection mechanisms such as process isolation introduces substantial security risks.
A single memory safety vulnerability~\cite{microsoft,google} can result in a full compromise of the entire cloud computing system.
Consequently, attackers can compromise other tenant's highly sensitive data.

Security vulnerabilities, such as Heartbleed~\cite{DBLP:conf/imc/DurumericKAHBLWABPP14} and Cloudbleed~\cite{cloudbleed1, cloudbleed2}, exemplify the severity of this attack surface.
In particular, Heartbleed allowed a remote adversary to leak private data (\eg confidential key material) by exploiting a buffer-overread error due to improper input sanitization triggered via maliciously crafted network packets. %
Without the strong isolation of memory resources, exploitable memory safety errors enable the leakage or corruption of sensitive process memory, such as private keys or authentication tokens. %
To address this security threat while preserving the level of optimization, cloud service providers are replacing process isolation with in-process sandboxes~\cite{CFModel}, \eg by using sandboxed languages like JavaScript~\cite{v8sandbox}, to efficiently support a high number of tenants per machine.
It is evident that cloud service providers have a strong need for scalable and efficient in-process isolation mechanisms that segregate memory resources of cloud-co-located tenants.

Common isolation techniques harden software systems by applying code instrumentation for logical protection with dynamic runtime checks.
Software-based fault isolation~(SFI)~\cite{DBLP:conf/sosp/WahbeLAG93,DBLP:journals/ftsec/Tan17} is an effective technique that instruments memory and control-flow operations to keep them within predefined regions.
Language-level isolation relies on partitioning the virtual address space to separate process memory~\cite{v8sandbox,v8sandbox2}.
This procedure enables software sandboxing, effectively restricting access for the sandbox to its assigned memory region.
Due to its efficiency, SFI has also been used in practice, as seen in the Google Native Client~(NaCl)~\cite{DBLP:conf/sp/YeeSDCMOONF09, DBLP:conf/uss/SehrMBKPSYC10} sandbox.
However, SFI can only offer coarse-grained sandboxing by isolating predetermined and contiguous memory regions without interleaving.
This means that SFI lacks fine-grained, object-level access control and the ability to dynamically manage resources in memory.

The efficiency of isolation can be significantly enhanced with hardware support.
Hardware-based isolation mechanisms, such as protection keys for user space~(PKU)~\cite{DBLP:journals/ieeesp/ParkLK23}, enable the restriction of access to memory at a page granularity.
Specifically, memory protection keys~(MPK) use a 4-bit protection key located in the page table entries~(PTE) to assign and enforce access policies for the respective page.
These policies are software-controlled via a processor register, associating two bits for every protection key to dynamically control read and write permissions. %
Thereby, MPK allows the partitioning of the process memory into 16 distinct regions, often called domains, and selectively controls their access.
Various PKU-based protection schemes~\cite{DBLP:conf/uss/Vahldiek-Oberwagner19, DBLP:conf/uss/SchrammelWSS0MG20,DBLP:conf/uss/SchrammelWSM22, DBLP:conf/usenix/HedayatiGJCSSM19,DBLP:conf/asiaccs/BlairRE23} are proposed to establish lightweight and transparent in-process isolation that is efficiently enforced in hardware. %
However, MPK is limited by its 4-bit key size, only offering 16 distinct protection domains.
This is restrictive for software systems that may require thousands of concurrent domains, \eg Cloudflare Workers~\cite{CFWorkers,CFModel}, within a single process.
Also, MPK is bound to the page granularity and cannot provide sub-page granular isolation.

In-process isolation is a highly demanded security feature for modern computing systems, such as cloud computing, imposing hard requirements on sandboxing techniques.
The sandbox must provide scalable isolation with fine-grained access control applicable to individual objects and memory pages.
Additionally, the technique must support thousands of concurrent sandboxes, \eg to isolate cloud workers, and must be available on commodity x86 server-class processors.

\paragraph{Contributions}
In this paper, we present \tmebox, a novel sandboxing technique for scalable in-process isolation that repurposes Intel's newly introduced total memory encryption multi-key~(\tmemk)~\cite{intelmktme} technology. %
\tmemk's runtime memory encryption is essentially used by the Intel trusted domain extensions~(TDX)~\cite{DBLP:journals/csur/ChengOVAGJFB24,inteltdxxwhitepaper}, their confidential computing architecture, to ensure the confidentiality and integrity of memory resources. %
This encryption and other architectural elements (\eg secure extended page tables~(EPT)~\cite{intelguide-3a}) allow TDX to provide strong isolation of virtual machines~(VMs), which run isolated workloads on the same physical machine. %

Intel \tmemk provides page-granular encryption of the computer's physical memory, which is primarily designed for the cryptographic isolation of VMs.
This enables hypervisors to encrypt VMs and containers~\cite{intelmktmeruntimeencryption}, as well as to protect against physical attacks on the memory subsystem~\cite{inteltdxxwhitepaper, DBLP:conf/uss/HaldermanSHCPCFAF08}.
However, the heavy-weight VM isolation of TDX incurs significant performance overhead, and \tmemk's encryption has not yet been applied for generic and efficient in-process memory isolation comparable to MPK.

\tmebox extends the application of the \tmemk memory encryption beyond hardware-isolated VMs by providing fine-grained and scalable in-process isolation on commodity x86 CPUs.
\tmebox deploys compiler instrumentation to isolate the code and data of mutually untrusted sandboxes.
Specifically, we enforce the usage of sandbox-specific \tmemk encryption keys by controlling the base address and index of memory operations.
Thereby, memory resources of the in-process sandbox are cryptographically isolated, leading to the detection of unauthorized memory access through hardware-backed integrity protection.
Repurposing the \tmemk memory encryption for software sandboxing offers several advantages over established in-process isolation mechanisms and addresses the hard requirements of modern computing systems, such as cloud computing.

First, \tmebox enforces fine-grained access control of memory resources.
Through the use of page aliasing, \tmebox achieves sub-page granular encryption and, thus, fine-grained memory isolation. %
Particularly, our design allows for scalable isolation granularities ranging from single cache lines to full pages, especially relevant in modern cloud settings.
In contrast, SFI-based mechanisms can only isolate coarse-grained, predetermined, and contiguous memory regions, restricting access by partitioning the virtual address space.

Second, \tmebox supports the flexible relocation of data in memory, which is particularly important for data-centric computation. %
Data relocation is essential in the cloud, \eg when allocator caches with smaller heap slots are used.
This enables flexible memory management for the allocator, providing continuous memory utilization across different sandboxes and efficient memory migration.

Third, our \tmebox design leverages a larger amount of key identifiers~(\keyids), enabling the support of thousands of concurrent sandboxes.
Compared to MPK's 4-bit protection keys (enabling 16 distinct isolation domains), our design leverages \tmemk that supports up to 15-bit \keyids, addressing up to 32K encryption keys~\cite{intelmktme} that we repurpose to cryptographically isolate sandboxes.
This is particularly relevant for software systems that require a large number of sandboxes, \eg cloud workers~\cite{CFWorkers,CFModel}, within a single process.
Additionally, our design supports frequent policy changes from user space without kernel interaction. %

Complementary to our design, we detail our prototype implementation of the \tmebox framework, consisting of an LLVM compiler toolchain and a security-hardened allocator.
Furthermore, we outline architecture-specific optimizations, such as x86 segment-based addressing~\cite{segementregister}, to achieve practical performance results for SPEC CPU2017~\cite{DBLP:conf/wosp/BucekLK18} and NGINX.
We showcase a geomean overhead of \SPECgs for data isolation and \SPECgscfi for code and data isolation using SPEC CPU2017.

In summary, the main contributions of this work are:

\begin{compactitem}
  \item \textbf{\tmebox.} We are the first to present a novel in-process isolation technique by repurposing the Intel \tmemk memory encryption on commodity x86 CPUs.
  \item \textbf{New Insights on Sandboxing.} We provide new insights on hardware-assisted sandboxing through memory encryption for modern cloud settings, enabling fine-grained and scalable isolation from single cache lines to full pages while supporting up to 32K sandboxes.  %
  \item \textbf{Prototype and Evaluation.} We implement an optimized prototype of \tmebox, utilizing x86 segment-based addressing, that showcases competitive performance results for SPEC CPU2017 and NGINX.
  \item \textbf{Security Analysis.} We systematically analyze the security threats and outline the derived security properties of our sandbox design.
\end{compactitem}

\paragraph{Outline}
The paper is structured as follows.
\Cref{sec:tmebox:background} provides the background of this work.
\Cref{sec:tmebox:threat_model} defines our threat model.
\Cref{sec:tmebox:design} and \Cref{sec:tmebox:implementation} describe \tmebox's design and implementation. %
\Cref{sec:tmebox:security_analysis} and \Cref{sec:tmebox:evaluation} provide the security analysis and evaluation.
\Cref{sec:tmebox:discussion} discusses related work, and \Cref{sec:tmebox:conclusion} concludes this paper.

\section{Background}\label{sec:tmebox:background}

This section provides the background on address translation with virtual memory, software-based fault isolation~(SFI), and Intel defensive execution technologies.

\subsection{Address Translation}
In modern operating systems, virtual memory organizes memory into contiguous blocks known as \emph{memory pages}, which are typically \SI{4}{\kilo\byte} in size on the x86-64 architecture. %
Pages are managed in a hierarchical structure called \emph{page tables}, enabling each user space application to have a distinct virtual representation of the computer's physical memory.

Address translation in paging systems is performed by the memory management unit~(MMU), which is responsible for resolving the virtual-to-physical mappings.
When accessing memory, the CPU translates a virtual address to a physical address via a page table walk.
The MMU uses the corresponding page table entries~(PTE) that contain the physical page number~(PPN) (and associated access permissions) to resolve the virtual address.
Frequently used virtual-to-physical mappings are cached in the translation look-aside buffer~(TLB) to enhance system performance.

Paging provides process isolation through virtual memory as an abstraction of the computer's physical memory.
Access protection is enforced in hardware and managed by the operating system, thus preventing processes from illegally accessing other processes' memory.

Typically, modern processors provide support for 48-bit or 57-bit virtual address spaces used for the address translation.
The virtualization of the computer's physical memory also allows multiple virtual addresses to refer to the same physical memory.
This is called \emph{aliasing} and is often used for shared memory (\ie code and data) across different processes.

\subsection{Software-based Fault Isolation}

Memory safety vulnerabilities~\cite{DBLP:conf/sp/SzekeresPWS13} frequently occur in complex software written in memory-unsafe programming languages (\eg C, C++)~\cite{microsoft, google}.
This is critical for software security, as an adversary can exploit memory safety vulnerabilities to compromise the target system~\cite{nsa, googlesecurebydesign}.
For instance, out-of-bounds~(OOB) errors, such as buffer over-reads and over-writes, enable an adversary to illegally access resources in memory.
The exploitation of memory safety errors allows the attacker to leak or corrupt sensitive data.

To mitigate this security threat, software sandboxing is used.
Sandboxes aim to reduce the attack surface of memory errors by isolating individual parts of the software.
Software-based fault isolation~(SFI)~\cite{DBLP:conf/sosp/WahbeLAG93} is a defense technique that implements in-process isolation (also referred to as intra-process or sub-process isolation), separating memory resources for individual software components within a single process.
Each SFI sandbox consists of an isolated data region containing runtime data (\eg stack and heap memory) and an isolated code region where the code of the sandbox resides~\cite{DBLP:journals/ftsec/Tan17}.
Consequently, the SFI sandbox needs to restrict data access and control-flow transfers.
Thus, all memory operations (\ie read and write accesses) performed by the sandbox's code are only allowed to access data within this isolated data region.
In addition, control-flow transfers must remain within the sandbox's code region (or target call sites that correspond to trusted runtime calls)~\cite{DBLP:journals/ftsec/Tan17}.

Data protection is achieved in traditional SFI systems by either using address checking or address masking, which instruments all memory operations (typically performed by the compiler) and enforces memory references to stay inside the sandbox boundaries~\cite{DBLP:conf/sosp/WahbeLAG93, DBLP:conf/uss/McCamantM06}. %
Modern SFI approaches implement data protection via the introduction of guard zones~\cite{DBLP:conf/ccs/ZengTM11}.
Guard zones refer to unmapped memory regions where memory accesses result in a page fault, thereby detecting accesses outside the sandbox.
SFI systems using guard zones need to ensure that memory operations remain within the sandbox's data and guard region by controlling the base and index register of memory operations.
For instance, Google Native Client~(NaCl)~\cite{DBLP:conf/sp/YeeSDCMOONF09, DBLP:conf/uss/SehrMBKPSYC10} uses \SI{4}{\giga\byte} virtual memory regions surrounded by \SI{40}{\giga\byte} guard zones to provide software sandboxing.

Control-flow transfers are restricted by SFI, enforcing that the sandboxed code remains in the designated code region~\cite{DBLP:conf/sp/YeeSDCMOONF09, DBLP:conf/uss/SehrMBKPSYC10}.
This is achieved by instrumenting indirect control transfers and function returns. %
Therefore, SFI prevents memory safety errors from corrupting memory outside a vulnerable software component, protecting the remaining system from exploitation and reducing the attack surface in the scenario of memory attacks.
The concept of SFI has been revisited for modern CPU architectures, such as x86 and ARM, proposing architecture-specific optimizations~\cite{DBLP:conf/sp/YeeSDCMOONF09, DBLP:conf/uss/SehrMBKPSYC10, DBLP:conf/ccs/ZhouWCW14, DBLP:conf/usenix/FordC08,DBLP:conf/pldi/MorrisettTTTG12,DBLP:conf/asplos/Yedidia24}.

\subsection{Intel Defensive Execution Technologies}

The Intel x86 architecture offers platform-specific hardware features that can be used for runtime protection.

\paragraph{Intel Control-Flow Enforcement Technology}
Control-flow enforcement technology~(CET)~\cite{DBLP:conf/isca/ShanbhogueGS19} is a set of processor extensions designed to implement control-flow integrity~(CFI)~\cite{DBLP:conf/ccs/AbadiBEL05,DBLP:journals/csur/BurowCNLFBP17,DBLP:conf/uss/CarliniBPWG15} measures into the CPU hardware.
CFI can be loosely categorized into two types: forward-edge and backward-edge CFI, which aim to protect function pointers and return addresses, respectively.

In the context of forward-edge CFI, Intel CET integrates indirect branch tracking~(IBT), utilizing so-called landing pads identified by the \texttt{endbr64} instructions inserted into function entries by the compiler.
Valid indirect jump targets are limited to function entries, thus reducing the attack surface of control-flow hijacking attacks such as jump-oriented programming~(JOP)~\cite{DBLP:conf/ccs/BletschJFL11}.
In terms of backward-edge CFI, Intel CET integrates a hardware shadow stack feature, which applies protection for return addresses.
The shadow stack feature achieves this by copying the return address onto the shadow stack, which is inaccessible to an attacker when entering a function.
On function exit, the return address restored from the regular stack is compared against the one on the shadow stack.
Mismatches are detected in hardware, and the potential corruption of return addresses is mitigated.
Thereby, CET thwarts code-reuse attacks such as return-into-libc~\cite{DBLP:conf/ccs/Shacham07} and return-oriented programming~(ROP)~\cite{DBLP:conf/ccs/BuchananRSS08}.

\paragraph{Intel Memory Protection Keys} Memory protection keys~(MPK)~\cite{DBLP:journals/ieeesp/ParkLK23} are a hardware feature that enforces page-level memory protection and allows changing permissions without requiring page table modifications.

MPK adds a 4-bit protection key encoded into the page table entries~(PTEs) and extends the processor architecture with a user space protection key register~(PKRU). %
The PKRU register enables the enforcement of software-controlled read and write permissions of pages by associating two bits for every distinct protection key.
MPK performs logical integrity checks during the address translation in the MMU by comparing the protection key with the current access permissions defined by the PKRU register.
In this way, MPK allows partitioning process memory into 16 distinct memory regions, dynamically controlling access permissions.
The PKRU register is accessible in user mode, which increases the performance of policy changes since they do not require kernel interaction. %

While originally introduced to enable execute-only memory and the locking-away of secrets, MPK has also been used to facilitate in-process isolation of untrusted code~\cite{DBLP:conf/usenix/HedayatiGJCSSM19,DBLP:conf/uss/Vahldiek-Oberwagner19,DBLP:conf/uss/SchrammelWSM22}.
However, using MPK for isolation has some drawbacks.
For instance, MPK's protection is limited for certain software architectures due to the relatively small 4-bit protection key.
Particularly, this means that MPK can only offer 16 distinct domains, which is restrictive for applications that may require thousands of separate domains (\eg FaaS applications) to isolate the memory of a single process.
Moreover, MPK applies its protection at the page-level.
Thus, fine-grained (\ie sub-page granular) isolation cannot be provided.

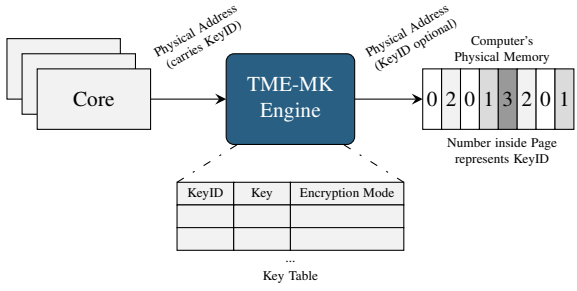
\begin{figure}
  \begin{center}
    \definecolor{fancyred}{HTML}{f70146}
\definecolor{fancypurple}{HTML}{6c2f91}
\definecolor{fancymid}{HTML}{5191c1}
\definecolor{fancygray}{HTML}{a5a5a5}
\definecolor{fancyblue}{HTML}{285f82}
\definecolor{fancygreen}{HTML}{78b473}
\definecolor{fancyyellow}{HTML}{ff8c00}
\definecolor{fancycyan}{HTML}{77babf}
\definecolor{fancyhead}{HTML}{245b78}
\definecolor{fancybody}{HTML}{e2e9ed}

\definecolor{flax}{HTML}{eedc82}
\definecolor{sand}{HTML}{c2b380}
\definecolor{fordtaupe}{HTML}{af9f96}
\definecolor{shared}{HTML}{ca9e92}
\definecolor{shareddark}{HTML}{a02060}

\newcommand\ptrY{-2.0}
\newcommand\ptrSize{0.4}

\newcommand\pagesize{1.5}
\newcommand\starta{0+\pagesize*1}
\newcommand\startb{0+\pagesize*1}
\newcommand\startc{0+\pagesize*3+0.1}
\newcommand\startd{0+\pagesize*4+0.1}
\newcommand\starte{0+\pagesize*5+0.2}
\newcommand\startf{0+\pagesize*2}
\newcommand\starti{0+\pagesize*2}
\newcommand\startg{0+\pagesize*2-\pagesize/2}
\newcommand\starth{0+\pagesize*3-\pagesize/2+0.1}

\begin{tikzpicture}
    \tikzset{edge/.style=->, >=stealth, thick};
    \tikzset{fmtsubfigure/.style={scale=0.65}};

    \draw[rounded corners=0pt, fill=fancygray!15] (0+\pagesize*0+0,-0.4+0.2) rectangle node[scale=0.8]{} ++(\pagesize,0.8);
    \draw[rounded corners=0pt, fill=fancygray!15] (0+\pagesize*0+0.2,-0.6+0.2) rectangle node[scale=0.8]{} ++(\pagesize,0.8);
    \draw[rounded corners=0pt, fill=fancygray!15] (0+\pagesize*0+0.4,-0.8+0.2) rectangle node[scale=0.8]{Core} ++(\pagesize,0.8);

    \draw[rounded corners=0pt, fill=fancygray!0] (0+\pagesize*4-0.5,-0.8+0.2) rectangle node[scale=0.8]{0} ++(0.25,0.8);
    \draw[rounded corners=0pt, fill=fancygray!15] (0+\pagesize*4-0.5+0.25*1,-0.8+0.2) rectangle node[scale=0.8]{2} ++(0.25,0.8);
    \draw[rounded corners=0pt, fill=fancygray!0] (0+\pagesize*4-0.5+0.25*2,-0.8+0.2) rectangle node[scale=0.8]{0} ++(0.25,0.8);
    \draw[rounded corners=0pt, fill=fancygray!40] (0+\pagesize*4-0.5+0.25*3,-0.8+0.2) rectangle node[scale=0.8]{1} ++(0.25,0.8);
    \draw[rounded corners=0pt, fill=black!40] (0+\pagesize*4-0.5+0.25*4,-0.8+0.2) rectangle node[scale=0.8]{3} ++(0.25,0.8);
    \draw[rounded corners=0pt, fill=fancygray!15] (0+\pagesize*4-0.5+0.25*5,-0.8+0.2) rectangle node[scale=0.8]{2} ++(0.25,0.8);
    \draw[rounded corners=0pt, fill=fancygray!0] (0+\pagesize*4-0.5+0.25*6,-0.8+0.2) rectangle node[scale=0.8]{0} ++(0.25,0.8);
    \draw[rounded corners=0pt, fill=fancygray!40] (0+\pagesize*4-0.5+0.25*7,-0.8+0.2) rectangle node[scale=0.8]{1} ++(0.25,0.8);

    \draw[rounded corners=3pt, fill=fancyblue] (0+\pagesize*2-0.1,-0.8) rectangle node[scale=0.8, text width=2cm, align=center, text=white]{TME-MK Engine} ++(\pagesize+0.2,1.2);

    \draw[rounded corners=0pt, draw=none] (0+\pagesize*4-0.4,-0.2) rectangle node[scale=0.5, text width=2.7cm, align=center]{Computer's Physical Memory} ++(\pagesize+0.4,1.3);
    \draw[rounded corners=0pt, draw=none] (0+\pagesize*4-0.4,-1.5) rectangle node[scale=0.5, text width=5cm, align=center]{Number inside Page represents KeyID} ++(\pagesize+0.4,1.2);

    \draw[rounded corners=4pt, draw=none] (0+\pagesize*1+0.5-0.3,0.2) rectangle node[scale=0.5,rotate around={30:(-1.4,0.8)}, text width=2.7cm, align=center]{Physical Address (carries KeyID)} ++(\pagesize,0);

    \draw[rounded corners=4pt, draw=none] (0+\pagesize*3+0.3-0.3,0.2) rectangle node[scale=0.5,rotate around={30:(-1.4,0.8)}, text width=2.7cm, align=center]{Physical Address (KeyID optional)} ++(\pagesize,0);

    \draw[rounded corners=3pt, draw=none] (\startg,\ptrY-0.7) rectangle node[scale=0.5]{Key Table} ++(\pagesize*2,0.3);
    \draw[rounded corners=3pt, draw=none] (\startg,\ptrY-0.5) rectangle node[scale=0.5]{...} ++(\pagesize*2,0.3);

    \draw[rounded corners=0pt, fill=fancygray!15] (\startg,-1.6) rectangle node[scale=0.5, text width=1cm, align=center]{KeyID} ++(0.75,0.3);
    \draw[rounded corners=0pt, fill=fancygray!15] (\startg+0.75,-1.6) rectangle node[scale=0.5, text width=1cm, align=center]{Key} ++(0.75,0.3);
    \draw[rounded corners=0pt, fill=fancygray!15] (\startg+0.75*2,-1.6) rectangle node[scale=0.5, text width=3cm, align=center]{Encryption Mode} ++(0.75*2,0.3);

    \draw[rounded corners=0pt, fill=fancygray!15] (\startg,-1.6-0.3) rectangle node[]{} ++(0.75,0.3);
    \draw[rounded corners=0pt, fill=fancygray!15] (\startg+0.75,-1.6-0.3) rectangle node[]{} ++(0.75,0.3);
    \draw[rounded corners=0pt, fill=fancygray!15] (\startg+0.75*2,-1.6-0.3) rectangle node[]{} ++(0.75*2,0.3);
    \draw[rounded corners=0pt, fill=fancygray!15] (\startg,-1.6-0.3*2) rectangle node[]{} ++(0.75,0.3);
    \draw[rounded corners=0pt, fill=fancygray!15] (\startg+0.75,-1.6-0.3*2) rectangle node[]{} ++(0.75,0.3);
    \draw[rounded corners=0pt, fill=fancygray!15] (\startg+0.75*2,-1.6-0.3*2) rectangle node[]{} ++(0.75*2,0.3);

    \draw[edge] (\pagesize+0.4,-0.2) -- ++(1,0);
    \draw[edge] (\pagesize*3+0.1,-0.2) -- ++(0.9,0);

    \draw[loosely dashdotted] (\startg,-1.3) -- ++(0.8,0.5);
    \draw[loosely dashdotted] (\starth+\pagesize-0.1,-1.3) -- ++(-0.8,0.5);

\end{tikzpicture}%
    \caption{High-level overview of the \tmemk encryption engine. The encryption engine processes the transferred data depending on the key identifier~(\keyid) used, which is encoded in the upper bits of the physical address. \tmemk uses a dedicated key table to resolve mappings from \keyids to encryption keys and encryption modes.}
    \label{fig:tmebox:background:tmemk_engine}
  \end{center}
\end{figure}

\paragraph{Intel Total Memory Encryption Multi-Key}
Total memory encryption~(TME)~\cite{intelmktme} is Intel's memory encryption technology that enables transparent encryption of DRAM data with a single encryption key.

The total memory encryption multi-key~(TME-MK)\footnote{Intel Total Memory Encryption Multi-Key~(TME-MK)~\cite{intelmktme} extension was introduced with the 3rd generation of Intel Xeon Scalable server processors.} extension enhances Intel TME with support for multiple encryption keys.
In general, memory encryption is a widely used technology that can provide confidentiality and integrity of DRAM memory.
The TME-MK feature is currently used for the cryptographic isolation of virtual machines~\cite{intelmktmeruntimeencryption} and for protection against physical attacks, such as cold boot attacks~\cite{inteltdxxwhitepaper, DBLP:conf/uss/HaldermanSHCPCFAF08}.

\Cref{fig:tmebox:background:tmemk_engine} provides a high-level overview of the Intel \tmemk encryption engine.
The \tmemk engine is located between the CPU core and the memory controller that interacts with the DRAM memory.
Moreover, \tmemk allows the use of up to $2^{15}$ keys, and it is platform-dependent how many are actually implemented.
The encryption engine maintains a key table that manages mappings of different key identifiers~(\keyids) that correspond to encryption keys (and encryption modes). Memory operations have a specific \keyid encoded into the upper part of the request's physical address, allowing the encryption of memory pages with different keys.
Note that the physical address and \keyid of the memory page are stored in the corresponding page table entry~(PTE) managed by the operating system.

\begin{figure}
  \begin{center}
    \definecolor{fancyred}{HTML}{f70146}
\definecolor{fancypurple}{HTML}{6c2f91}
\definecolor{fancymid}{HTML}{5191c1}
\definecolor{fancygray}{HTML}{a5a5a5}
\definecolor{fancyblue}{HTML}{285f82}
\definecolor{fancygreen}{HTML}{78b473}
\definecolor{fancyyellow}{HTML}{ff8c00}
\definecolor{fancycyan}{HTML}{77babf}
\definecolor{fancyhead}{HTML}{245b78}
\definecolor{fancybody}{HTML}{e2e9ed}

\definecolor{flax}{HTML}{eedc82}
\definecolor{sand}{HTML}{c2b380}
\definecolor{fordtaupe}{HTML}{af9f96}
\definecolor{shared}{HTML}{ca9e92}
\definecolor{shareddark}{HTML}{a02060}

\newcommand\ptrY{-2.0}
\newcommand\ptrSize{0.4}

\newcommand\pagesize{1.5}
\newcommand\starta{0+\pagesize*1}
\newcommand\startb{0+\pagesize*1}
\newcommand\startc{0+\pagesize*3+0.1}
\newcommand\startd{0+\pagesize*4+0.1}
\newcommand\starte{0+\pagesize*5+0.2}
\newcommand\startf{0+\pagesize*2}
\newcommand\starti{0+\pagesize*2}
\newcommand\startg{0+\pagesize*2-\pagesize/2}
\newcommand\starth{0+\pagesize*3-\pagesize/2+0.1}

\begin{tikzpicture}
    \tikzset{edge/.style=->, >=stealth, thick};
    \tikzset{fmtsubfigure/.style={scale=0.65}};

    \draw[rounded corners=0pt, dotted] (0+\pagesize*0,-0.8) rectangle node[scale=0.8]{} ++(\pagesize*5,1.2);
    \draw[rounded corners=0pt, dotted] (0+\pagesize*0,-2.6) rectangle node[scale=0.5]{} ++(\pagesize*5,0.4);

    \draw[rounded corners=0pt, draw=none] (0+\pagesize*0,-0.8) rectangle node[scale=0.5, text width=2.7cm, align=center]{Memory Controller} ++(\pagesize,1.2);
    \draw[rounded corners=0pt, draw=none] (0+\pagesize*0,-2.6) rectangle node[scale=0.5, text width=2.7cm, align=center]{DRAM} ++(\pagesize,0.4);
    \draw[rounded corners=0pt, draw=none] (0+\pagesize*0,-1.6) rectangle node[scale=0.5, text width=2.0cm, align=center]{TME-MK Engine} ++(\pagesize,0.4);

    \draw[rounded corners=3pt, fill=fancyblue] (\pagesize*1.5-0.4,-1.6) rectangle node[scale=0.6, text=white]{$E_K$} ++(\pagesize*0.5,0.5);

    \draw[rounded corners=3pt, fill=fancyblue] (\pagesize*2,-1.95) rectangle node[scale=0.6, text=white]{$H_K$} ++(\pagesize*0.5,0.4);

    \draw[rounded corners=3pt, fill=fancyblue] (\pagesize*3.25-0.4,-1.6) rectangle node[scale=0.6, text=white]{$D_K$} ++(\pagesize*0.5,0.5);
    \draw[rounded corners=3pt, fill=fancyblue] (\pagesize*3.5,-1.95) rectangle node[scale=0.6, text=white]{$H_K$} ++(\pagesize*0.5,0.4);

    \draw[rounded corners=0pt, fill=fancygray!15] (\pagesize,-0.5) rectangle node[scale=0.6]{Data} ++(\pagesize*1,0.4);

    \draw[rounded corners=0pt, fill=fancygray!15] (\pagesize,-2.6) rectangle node[scale=0.8]{} ++(\pagesize*1.5,0.4);
    \draw[rounded corners=0pt, fill=fancygray!15] (\pagesize,-2.6) rectangle node[scale=0.6]{Enc. Data} ++(\pagesize*1,0.4);
    \draw[rounded corners=0pt, fill=fancygray!15] (\pagesize*2,-2.6) rectangle node[scale=0.6]{MAC} ++(\pagesize*0.5,0.4);

    \draw[rounded corners=0pt, fill=fancygray!15] (\pagesize*3,-0.5) rectangle node[scale=0.6]{Data} ++(\pagesize*1,0.4);

    \draw[rounded corners=0pt, fill=fancygray!15] (\pagesize*3,-2.6) rectangle node[scale=0.8]{} ++(\pagesize*1.5,0.4);
    \draw[rounded corners=0pt, fill=fancygray!15] (\pagesize*3,-2.6) rectangle node[scale=0.6]{Enc. Data} ++(\pagesize*1,0.4);
    \draw[rounded corners=0pt, fill=fancygray!15] (\pagesize*3+\pagesize,-2.6) rectangle node[scale=0.6]{MAC} ++(\pagesize*0.5,0.4);

    \draw[rounded corners=0pt,draw=none] (\pagesize*3+\pagesize,-1.95) rectangle node[scale=0.6]{$\stackrel{?}{=}$} ++(\pagesize*0.5,0.4);
    \draw[rounded corners=0pt,draw=none] (\pagesize*3+\pagesize,-1.0) rectangle node[scale=0.6, text width=1.5cm, align=center]{Exception} ++(\pagesize*0.5,0.1);
    \draw[rounded corners=0pt,draw=none] (\pagesize*3+\pagesize,-1.25) rectangle node[scale=0.6, text width=1.5cm, align=center]{\faWarning} ++(\pagesize*0.5,0.1);

    \draw[edge] (\pagesize*3.5,-0.1) -- ++(0,0.25);
    \draw[edge] (\pagesize*3.25,-2.2) -- ++(0,0.6);
    \draw[edge] (\pagesize*3.75,-2.2) -- ++(0,0.25);
    \draw[edge] (\pagesize*3.25,-1.1) -- ++(0,0.6);
    \draw[edge] (\pagesize*4.25,-1.95-0.25) -- ++(0,0.25);
    \draw[edge] (\pagesize*4.25,-1.6) -- ++(0,0.25);
    \draw[edge] (\pagesize*4,-1.75) -- ++(0.25,0);

    \draw[edge] (\pagesize*1.5,0.15) -- ++(0,-0.25);
    \draw[edge] (\pagesize*1.5,-0.5) -- ++(0,-0.6);
    \draw[edge] (\pagesize*1.5,-1.6) -- ++(0,-0.6);
    \draw[edge] (\pagesize*2.25,-1.95) -- ++(0,-0.25);
    \draw[edge] (\pagesize*1.5,-1.6) -- ++(0,-0.15) -- ++(0.75,0);
    \draw[rounded corners=4pt, draw=none] (\pagesize*1.5,0.25) rectangle node[scale=0.6]{From CPU} ++(0,0); %
    \draw[rounded corners=4pt, draw=none] (\pagesize*3.5,0.25) rectangle node[scale=0.6]{To CPU} ++(0,0); %

\end{tikzpicture}%
    \begin{subfigure}[b]{0.2\textwidth}\vspace*{2mm}\caption{Memory Encryption}\label{fig:tmebox:background:tmemk_encryption:a}\end{subfigure}
    \begin{subfigure}[b]{0.2\textwidth}\vspace*{2mm}\caption{Memory Decryption}\label{fig:tmebox:background:tmemk_encryption:b}\end{subfigure}
    \caption{Overview of \tmemk's memory encryption with integrity support. Memory operations leverage a specific encryption key that is determined by the \keyid to encrypt and decrypt data located in DRAM. Memory accesses using an incorrect \keyid are detected by a MAC mismatch and result in a hardware exception.}
    \label{fig:tmebox:background:tmemk_encryption}
  \end{center}
\end{figure}
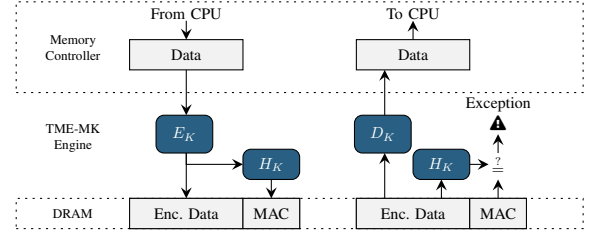

\Cref{fig:tmebox:background:tmemk_encryption} illustrates the encryption procedure of \tmemk with integrity support.
The \tmemk encryption engine uses AES~\cite{DBLP:books/sp/DaemenR02, DBLP:conf/cardis/DaemenR98a} in XTS mode~\cite{DBLP:journals/ieeesp/Martin10, DBLP:conf/ccs/RogawayBBK01}, supporting both 128-bit and 256-bit encryption keys.
In addition, with the introduction of the Intel trust domain extensions~(TDX)~\cite{DBLP:journals/csur/ChengOVAGJFB24,inteltdxxwhitepaper}, authenticated encryption is integrated.
This extends TME-MK with support for data integrity using a message authentication code~(MAC). %
The MAC is computed using the SHA-3~\cite{guido2011k} secure hash algorithm.
Memory operations are performed using the associated \keyid that selects the used encryption key.
When writing to memory, TME-MK encrypts the data and computes the MAC to provide confidentiality and integrity of DRAM data (\cf \Cref{fig:tmebox:background:tmemk_encryption:a} - Memory Encryption).
When reading encrypted data from DRAM, it decrypts and authenticates the data by recomputing and comparing the associated MAC.
An integrity violation is detected within the cryptographic bounds of the MAC (\cf \Cref{fig:tmebox:background:tmemk_encryption:b} - Memory Decryption).
In case of an integrity violation, either a fixed pattern is returned to prevent ciphertext analysis~\cite{inteltdxxwhitepaper}, or a hardware exception is raised~\cite{DBLP:conf/eurosp/SchrammelULSGLDM24}.

TME-MK also allows the configuration of the used encryption mode, \ie the key size, and to enable integrity support.
When using integrity, each MAC covers one cache line. %
This means that the smallest usable granularity for encryption with integrity mode is also cache line-sized (\ie \SI{64}{\byte}).

\section{Threat Model}\label{sec:tmebox:threat_model}

We consider a strong adversary that intends to exploit one or several memory safety vulnerabilities~\cite{DBLP:conf/sp/SzekeresPWS13} present in the target unprivileged user space program, thereby gaining unauthorized access to resources in memory. %
The attacker obtains an arbitrary (read and write) memory access primitive, \eg by triggering a buffer over-read or over-write error via maliciously crafted user input.
They then exploit this vulnerability in an attempt to leak or corrupt sensitive process memory (\eg confidential data such as private keys or authentication tokens).

Also, we assume that the adversary possesses knowledge of the process's address space layout (\ie the attacker can circumvent ASLR~\cite{DBLP:conf/secdev/GanzP17, DBLP:conf/ccs/ShachamPPGMB04}).
However, we assume that privileged software (\ie the operating system or hypervisor) is benign and free of exploitable programming errors.
Moreover, we presume that Write-XOR-Execute is enabled by default.
Thus, the attacker cannot perform code-injection attacks~\cite{DBLP:journals/ieeesp/Arce04b} due to the enforcement of the no-execution policy on writable memory.
Side channels~\cite{DBLP:conf/sp/KocherHFGGHHLM019, DBLP:conf/uss/Lipp0G0HFHMKGYH18} and fault injection attacks~\cite{DBLP:conf/isca/KimDKFLLWLM14, DBLP:conf/sp/MurdockOGBGP20} are considered out-of-scope for this work.

\section{\tmebox System Design}\label{sec:tmebox:design}

This section presents \tmebox, a novel technique for strong and efficient in-process isolation built on top of the hardware-backed \tmemk encryption.
\tmebox repurposes Intel \tmemk, intended to encrypt the memory of virtual machines, for fine-grained and scalable in-process isolation on commodity Intel x86 CPUs.
By applying compiler instrumentation, \tmebox enforces that sandboxes use a designated encryption key for memory interactions.
Thereby, \tmebox cryptographically isolates memory, allowing the detection of unauthorized accesses from mutually untrusted sandboxes.

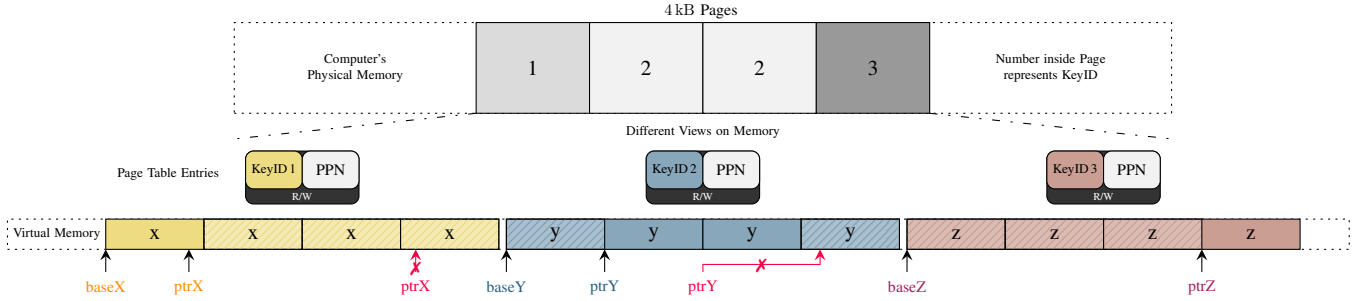
\begin{figure*}
  \captionsetup[subfigure]{font=scriptsize}
  \begin{center}
    \definecolor{fancyred}{HTML}{f70146}
\definecolor{fancypurple}{HTML}{6c2f91}
\definecolor{fancymid}{HTML}{5191c1}
\definecolor{fancygray}{HTML}{a5a5a5}
\definecolor{fancyblue}{HTML}{285f82}
\definecolor{fancygreen}{HTML}{78b473}
\definecolor{fancyyellow}{HTML}{ff8c00}
\definecolor{fancycyan}{HTML}{77babf}
\definecolor{fancyhead}{HTML}{245b78}
\definecolor{fancybody}{HTML}{e2e9ed}

\definecolor{flax}{HTML}{eedc82}
\definecolor{sand}{HTML}{c2b380}
\definecolor{fordtaupe}{HTML}{af9f96}
\definecolor{shared}{HTML}{ca9e92}
\definecolor{shareddark}{HTML}{a02060}

\newcommand\ptrY{-2.0}
\newcommand\ptrSize{0.4}

\newcommand\pagesize{1.5}
\newcommand\domainsize{1.3}
\newcommand\starta{0+\domainsize*1}
\newcommand\startb{0+\domainsize*1}
\newcommand\startc{0+\domainsize*5+0.1}
\newcommand\startd{0+\domainsize*7+0.1}
\newcommand\starte{0+\domainsize*9+0.2}
\newcommand\startf{0+\pagesize*8}
\newcommand\starti{0+\pagesize*9}
\newcommand\startg{0+\pagesize*9-\pagesize/2}
\newcommand\starth{0+\pagesize*10-\pagesize/2+0.1}

\begin{tikzpicture}
    \tikzset{edge/.style=->, >=stealth, thick};
    \tikzset{fmtsubfigure/.style={scale=0.65}};

    \draw[rounded corners=0pt, dotted] (0+\pagesize*2,-0.8) rectangle node[scale=0.8]{} ++(\pagesize*11-4.1,1.2);
    \draw[rounded corners=0pt, dotted] (0+\pagesize*0,-2.6) rectangle node[scale=0.5]{} ++(\pagesize*11+1.25,0.4);

    \draw[rounded corners=0pt, fill=fancygray!40] (0+\pagesize*4+0.2,-0.8) rectangle node[scale=0.8]{1} ++(\pagesize,1.2);
    \draw[rounded corners=0pt, fill=fancygray!15] (0+\pagesize*5+0.2,-0.8) rectangle node[scale=0.8]{2} ++(\pagesize,1.2);
    \draw[rounded corners=0pt, fill=fancygray!15] (0+\pagesize*6+0.2,-0.8) rectangle node[scale=0.8]{2} ++(\pagesize,1.2);
    \draw[rounded corners=0pt, fill=black!40] (0+\pagesize*7+0.2,-0.8) rectangle node[scale=0.8]{3} ++(\pagesize,1.2);

    \draw[rounded corners=0pt, draw=none] (0+\pagesize*2,-0.8) rectangle node[scale=0.5, text width=2.7cm, align=center]{Computer's Physical Memory} ++(\pagesize*2+0.2,1.2);
    \draw[rounded corners=0pt, draw=none] (0+\pagesize*8+0.2,-0.8) rectangle node[scale=0.5, text width=3.5cm, align=center]{Number inside Page represents KeyID} ++(\pagesize*2+0.2,1.2);
    \draw[rounded corners=0pt, draw=none] (0+\pagesize*0-0.1,-2.6) rectangle node[scale=0.5, text width=2.7cm, align=center]{Virtual Memory} ++(\pagesize,0.4);
    \draw[rounded corners=3pt, draw=none] (1+\pagesize*0.25,\ptrY) rectangle node[scale=0.5, text width=4.3cm, align=center]{Page Table Entries} ++(\pagesize,0.8);

    \draw[rounded corners=4pt, draw=none] (\startb+\pagesize*3.25,0.55) rectangle node[scale=0.6]{\SI{4}{\kilo\byte} Pages} ++(\pagesize*4,0);

    \draw[rounded corners=3pt, fill=black!80] (\startb+\domainsize*2-\pagesize/2,\ptrY) rectangle node[scale=0.7]{} ++(\pagesize,0.7);
    \draw[rounded corners=3pt, draw=none] (\startb+\domainsize*2-\pagesize/2,\ptrY) rectangle node[scale=0.4,text=white]{R/W} ++(\pagesize,0.2);
    \draw[rounded corners=3pt, fill=flax] (\startb+\domainsize*2-\pagesize/2,-1.8) rectangle node[scale=0.5, text width=1.4cm, align=center]{KeyID\,1} ++(0.75,0.5);
    \draw[rounded corners=3pt, fill=fancygray!15] (\startb+\domainsize*2-\pagesize/2+0.75,-1.8) rectangle node[scale=0.6, text width=1cm, align=center]{PPN} ++(0.75,0.5);

    \draw[rounded corners=0pt, fill=flax] (\startb,-2.6) rectangle node[scale=0.8]{x} ++(\domainsize,0.4);
    \draw[rounded corners=0pt, fill=flax!60] (\startb+\domainsize,-2.6) rectangle node[scale=0.8]{} ++(\domainsize,0.4);
    \draw[rounded corners=0pt, pattern=north east lines, pattern color=flax] (\startb+\domainsize,-2.6) rectangle node[scale=0.8]{x} ++(\domainsize,0.4);
    \draw[rounded corners=0pt, fill=flax!60] (\startb+\domainsize*2,-2.6) rectangle node[scale=0.8]{} ++(\domainsize,0.4);
    \draw[rounded corners=0pt, pattern=north east lines, pattern color=flax] (\startb+\domainsize*2,-2.6) rectangle node[scale=0.8]{x} ++(\domainsize,0.4);
    \draw[rounded corners=0pt, fill=flax!60] (\startb+\domainsize*3,-2.6) rectangle node[scale=0.8]{} ++(\domainsize,0.4);
    \draw[rounded corners=0pt, pattern=north east lines, pattern color=flax] (\startb+\domainsize*3,-2.6) rectangle node[scale=0.8]{x} ++(\domainsize,0.4);

    \draw[rounded corners=3pt, fill=black!80] (\startc+\domainsize*2-\pagesize/2,\ptrY) rectangle node[scale=0.7]{} ++(\pagesize,0.7);
    \draw[rounded corners=3pt, draw=none] (\startc+\domainsize*2-\pagesize/2,\ptrY) rectangle node[scale=0.4,text=white]{R/W} ++(\pagesize,0.2);
    \draw[rounded corners=3pt, fill=fancyblue!55] (\startc+\domainsize*2-\pagesize/2,-1.8) rectangle node[scale=0.5, text width=1.4cm, align=center]{KeyID\,2} ++(0.75,0.5);
    \draw[rounded corners=3pt, fill=fancygray!15] (\startc+\domainsize*2-\pagesize/2+0.75,-1.8) rectangle node[scale=0.6, text width=1cm, align=center]{PPN} ++(0.75,0.5);

    \draw[rounded corners=0pt, fill=fancyblue!40] (\startc,-2.6) rectangle node[scale=0.8]{} ++(\domainsize,0.4);
    \draw[rounded corners=0pt, pattern=north east lines, pattern color=fancyblue!60] (\startc,-2.6) rectangle node[scale=0.8]{y} ++(\domainsize,0.4);
    \draw[rounded corners=0pt, fill=fancyblue!55] (\startc+\domainsize,-2.6) rectangle node[scale=0.8]{y} ++(\domainsize,0.4);

    \draw[rounded corners=0pt, fill=fancyblue!55] (\startd,-2.6) rectangle node[scale=0.8]{y} ++(\domainsize,0.4);
    \draw[rounded corners=0pt, fill=fancyblue!40] (\startd+\domainsize,-2.6) rectangle node[scale=0.8]{} ++(\domainsize,0.4);
    \draw[rounded corners=0pt, pattern=north east lines, pattern color=fancyblue!60] (\startd+\domainsize,-2.6) rectangle node[scale=0.8]{y} ++(\domainsize,0.4);

    \draw[rounded corners=3pt, fill=black!80] (\starte+\domainsize*2-\pagesize/2,\ptrY) rectangle node[scale=0.7]{} ++(\pagesize,0.7);
    \draw[rounded corners=3pt, draw=none] (\starte+\domainsize*2-\pagesize/2,\ptrY) rectangle node[scale=0.4,text=white]{R/W} ++(\pagesize,0.2);
    \draw[rounded corners=3pt, fill=shared] (\starte+\domainsize*2-\pagesize/2,-1.8) rectangle node[scale=0.5, text width=1.4cm, align=center]{KeyID\,3} ++(0.75,0.5);
    \draw[rounded corners=3pt, fill=fancygray!15] (\starte+\domainsize*2-\pagesize/2+0.75,-1.8) rectangle node[scale=0.6, text width=1cm, align=center]{PPN} ++(0.75,0.5);

    \draw[rounded corners=0pt, fill=shared!70] (\starte,-2.6) rectangle node[scale=0.8]{} ++(\domainsize,0.4);
    \draw[rounded corners=0pt, pattern=north east lines, pattern color=shared] (\starte,-2.6) rectangle node[scale=0.8]{z} ++(\domainsize,0.4);
    \draw[rounded corners=0pt, fill=shared!70] (\starte+\domainsize,-2.6) rectangle node[scale=0.8]{} ++(\domainsize,0.4);
    \draw[rounded corners=0pt, pattern=north east lines, pattern color=shared] (\starte+\domainsize,-2.6) rectangle node[scale=0.8]{z} ++(\domainsize,0.4);
    \draw[rounded corners=0pt, fill=shared!70] (\starte+\domainsize*2,-2.6) rectangle node[scale=0.8]{} ++(\domainsize,0.4);
    \draw[rounded corners=0pt, pattern=north east lines, pattern color=shared] (\starte+\domainsize*2,-2.6) rectangle node[scale=0.8]{z} ++(\domainsize,0.4);
    \draw[rounded corners=0pt, fill=shared] (\starte+\domainsize*3,-2.6) rectangle node[scale=0.8]{z} ++(\domainsize,0.4);

    \draw[edge, fancyred] (\starta+\domainsize*3+0.2,-2.9) -- ++(0,0.3) node[midway,yshift=-0.3em]{\scriptsize\ding{55}};
    \draw[rounded corners=4pt, draw=none] (\starta+\domainsize*3+0.2,-3.1) rectangle node[scale=0.6]{{\color{fancyred}ptrX}} ++(0,0);

    \draw[edge] (\starta+\domainsize*1-0.2,-2.9) -- ++(0,0.3) node[midway,yshift=-0.3em]{};
    \draw[rounded corners=4pt, draw=none] (\starta+\domainsize*1-0.2,-3.1) rectangle node[scale=0.6]{{\color{fancyyellow}ptrX}} ++(0,0);

    \draw[edge] (\starta + \domainsize*5+0.1,-2.9) -- ++(0,0.3) node[midway,yshift=-0.3em]{};
    \draw[rounded corners=4pt, draw=none] (\starta + \domainsize*5+0.1,-3.1) rectangle node[scale=0.6]{{\color{fancyblue}ptrY}} ++(0,0);

    \draw[edge] (\startb,-2.9) -- ++(0,0.3);
    \draw[rounded corners=4pt, draw=none] (\startb,-3.1) rectangle node[scale=0.6]{{\color{fancyyellow}baseX}} ++(0,0);

    \draw[edge] (\startc,-2.9) -- ++(0,0.3);
    \draw[rounded corners=4pt, draw=none] (\startc,-3.1) rectangle node[scale=0.6]{{\color{fancyblue}baseY}} ++(0,0);

    \draw[edge, fancyred] (\startd,-2.9) -- ++(0,0.1) -- ++(1.55,0) node[midway,yshift=-0.05em]{\scriptsize\ding{55}} -- ++(0,0.2);
    \draw[rounded corners=4pt, draw=none] (\startd,-3.1) rectangle node[scale=0.6]{{\color{fancyred}ptrY}} ++(0,0);

    \draw[edge] (\starte,-2.9) -- ++(0,0.3);
    \draw[rounded corners=4pt, draw=none] (\starte,-3.1) rectangle node[scale=0.6]{{\color{shareddark}baseZ}} ++(0,0);

    \draw[edge] (\starte+\domainsize*3,-2.9) -- ++(0,0.3);
    \draw[rounded corners=4pt, draw=none] (\starte+\domainsize*3,-3.1) rectangle node[scale=0.6]{{\color{shareddark}ptrZ}} ++(0,0); %

    \draw[draw=none] (\startc+\domainsize*2-\pagesize/2,\ptrY+0.7) rectangle node[scale=0.5, text width=5.7cm, align=center]{Different Views on Memory} ++(\pagesize,0.5);
    \draw[loosely dashdotted] (\pagesize*4+0.2,-0.8) -- ++(-\pagesize*2-0.3,-0.35);
    \draw[loosely dashdotted] (\pagesize*8+0.2,-0.8) -- ++(\pagesize*2+0.3,-0.35);

\end{tikzpicture}%
    \caption{
      Overview of \tmebox's page granular isolation by illustrating the virtual address space of three different sandboxes that are mapped to four physical memory pages.
      Memory operations of the sandboxes \texttt{x}, \texttt{y}, and \texttt{z} use the \keyids 1, 2, and 3, respectively.
      The physical pages are encrypted with the encryption keys of their corresponding sandbox, restricting access solely to their \keyids.
      When a sandbox accesses pages that are not assigned to it, \ie unauthorized access of a non-adjacent or adjacent page, the violation is detected by the integrity checks of the \tmemk encryption engine.
    }
    \label{fig:tmebox:design:isolation}
  \end{center}
\end{figure*}

\subsection{Secure System Architecture}

\paragraph{Design Properties}
In the following, we outline the main design properties of \tmebox, addressing the hard requirements of modern computing systems, such as cloud computing.
We enable scalable isolation and flexible data relocation without introducing new software dependencies, \eg hard constraints for the memory allocator.
We provide cryptographic in-process isolation, thereby supporting a large number of sandboxes (\eg for isolating cloud workers) and detecting unauthorized access through integrity exceptions.
In this way, \tmebox enables hardware-supported in-process isolation on commodity x86 machines for modern cloud settings.

\begin{compactitem}
  \item \textbf{Scalable isolation:} \tmebox enforces fine-grained and scalable access control, supporting isolation granularities ranging from individual cache lines to full pages. %
  \item \textbf{Flexible data relocation:} \tmebox allows flexible relocation of data in memory, enabling efficient memory migration and continuous memory operation. %
  \item \textbf{Number of sandboxes:} \tmebox supports up to 32K sandboxes, as Intel \tmemk provides up to $2^{15}$ encryption keys that we repurpose for cryptographic isolation. %
  \item \textbf{Integrity enforcement:} The \tmemk encryption engine ensures data integrity for memory operations, allowing us to detect unauthorized sandbox accesses.
  \item \textbf{Commodity hardware:} \tmebox repurposes Intel \tmemk, a platform-specific hardware feature available on commodity x86-64 CPUs.
\end{compactitem}

\paragraph{Overview}
At its core, \tmebox repurposes the \tmemk encryption engine for efficient and secure in-process isolation by enforcing the usage of a sandbox-specific alias that represents a \tmemk key identifier~(\keyid).
Each sandbox is assigned its dedicated \keyid, mapping to the sandbox's encryption key.
Moreover, compiler-based code instrumentation ensures that all memory interactions of the sandbox use the sandbox-specific \keyid, thereby cryptographically isolating the memory of sandboxes from each other.
To achieve this, \tmebox stores the sandbox's base address in a separate CPU register, protecting it from malicious access.
This register can either be a general-purpose register or an x86 segment register~\cite{segementregister}.
\tmebox enforces memory isolation by controlling the base address and index of memory operations.

\tmebox leverages different views on the computer's physical memory to provide scalable isolation for sandboxes.
The base address is used to select a sandbox-specific \keyid through the associated memory alias, ensuring that memory accesses within the sandbox always use its respective encryption key.
If a sandbox attempts to access data from another sandbox with an incorrect \keyid, the underlying encryption triggers an exception, detecting unauthorized access.

Thereby, \tmebox achieves isolation granularities ranging from individual cache lines to full pages.
We combine flexible and software-controlled access policies with the cryptographic integrity enforcement of \tmemk{}. %
This means that we are not limited to contiguous memory ranges and can quickly re-assign individual small memory granules to different sandboxes.
Sandboxes can grow or shrink in size, down to the granularity of the underlying memory encryption, which is \SI{64}{\byte} cache line size.
\tmebox also allows the flexible relocation of memory depending on runtime constraints.
Moreover, we can provide numerous sandboxes, \ie \tmemk supports up to $2^{15}$ \keyids addressing 32K encryption keys that we use for isolation. %

\tmebox ensures the isolation of runtime data (\eg stack and heap memory) and restricts control-flow transfers of its code region.
During the startup of each sandbox, it initializes the sandbox's stack and static memory with the corresponding \keyid and sets up the base address register.
The virtual address space is mapped so that required memory regions, such as heap memory, are accessible through each sandbox's designated \keyid.
For this, we use memory aliasing with different \keyids so that multiple sandboxes can share the same underlying physical memory.
This approach provides scalable isolation granularities for dynamic memory and enables flexible resource management.
Additionally, \tmebox instruments control flow transfers to ensure that they reside within the sandbox's code region.
Our compiler instruments indirect function calls and jumps, \ie function pointers, by applying address masking.
Similarly, return addresses are instrumented to ensure control-flow transfers remain within the sandbox.

The memory allocator can dynamically adjust security policies for heap memory, \ie enable and revoke memory access. %
The \tmemk hardware performs integrity checks, providing the ability to easily manage and relocate data in memory.
Once the \keyids are set for the virtual address space, individual cache lines of memory can be assigned to a sandbox with a single memory write without kernel interaction.
This enables efficient resource management through the flexible relocation of data. %
Our approach enhances performance, especially when frequent policy changes occur, applying fast changes to the protected regions of the sandbox's memory.

\subsection{Scalable Memory Isolation and Flexible Data Relocation}\label{sec:tmebox:design:isolation}

In the following, we discuss \tmebox's scalable isolation and flexible data relocation in memory.
We first detail how entire memory pages (or ranges of pages) are isolated from each other.
Next, we elaborate on the fine-grained memory isolation and flexible relocation that \tmebox offers.

\paragraph{Scalable Memory Isolation}
\tmebox can provide isolation for full pages assigned to different sandboxes.
Our software design maps the virtual address space of the sandboxes to the same physical memory through page aliasing.
Furthermore, our design leverages compiler instrumentation to enforce that the memory operations of the sandbox always use its associated \keyid located in the PTE.
By controlling the base address of memory operations, we ensure that accesses are performed with the corresponding encryption key of the sandbox and validated by  \tmemk. %

Moreover, this allows the memory allocator to efficiently manage the memory for different sandboxes by initializing the distinct memory locations (\ie writing an entire cache line with a \keyid to initialize the MAC).
\tmebox supports the isolation of a large number of pages, including non-contiguous memory regions. %
\Cref{fig:tmebox:design:isolation} illustrates \tmebox's page-granular isolation, showing how different sandboxes can have distinct views of the computer's physical memory.

Specifically, the virtual address space of three distinct sandboxes (\texttt{x}, \texttt{y}, and \texttt{z}) is mapped to four pages of the computer's physical memory.
The sandboxes use the \keyids 1, 2, and 3, respectively, to cryptographically isolate their memory.
Each sandbox is restricted to accessing its own memory since the physical pages are encrypted with different encryption keys of the corresponding sandbox, \ie memory accesses are only granted by using the correct sandbox-specific encryption key.
Precisely, the example shows unauthorized access to a non-adjacent page by sandbox \texttt{x} and to an adjacent page by sandbox \texttt{y}.
Invalid accesses are detected by \tmemk's cryptographic integrity checks. %
This allows us to assign and isolate individual pages for different sandboxes.

\begin{figure}
  \begin{center}
    \definecolor{fancyred}{HTML}{f70146}
\definecolor{fancypurple}{HTML}{6c2f91}
\definecolor{fancymid}{HTML}{5191c1}
\definecolor{fancygray}{HTML}{a5a5a5}
\definecolor{fancyblue}{HTML}{285f82}
\definecolor{fancygreen}{HTML}{78b473}
\definecolor{fancyyellow}{HTML}{ff8c00}
\definecolor{fancycyan}{HTML}{77babf}
\definecolor{fancyhead}{HTML}{245b78}
\definecolor{fancybody}{HTML}{e2e9ed}

\definecolor{flax}{HTML}{eedc82}
\definecolor{sand}{HTML}{c2b380}
\definecolor{fordtaupe}{HTML}{af9f96}
\definecolor{shared}{HTML}{ca9e92}
\definecolor{shareddark}{HTML}{a02060}

\newcommand\ptrY{-2.0}
\newcommand\ptrSize{0.4}

\newcommand\pagesize{1.5}
\newcommand\starta{0+\pagesize*1}
\newcommand\startb{0+\pagesize*1}
\newcommand\startc{0+\pagesize*3+0.1}
\newcommand\startd{0+\pagesize*4+0.1}
\newcommand\starte{0+\pagesize*5+0.2}
\newcommand\startf{0+\pagesize*2}
\newcommand\starti{0+\pagesize*2}
\newcommand\startg{0+\pagesize*2-\pagesize/2}
\newcommand\starth{0+\pagesize*3-\pagesize/2+0.1}

\begin{tikzpicture}
    \tikzset{edge/.style=->, >=stealth, thick};
    \tikzset{fmtsubfigure/.style={scale=0.65}};

    \draw[rounded corners=0pt, dotted] (0+\pagesize*0+0.25,-0.8) rectangle node[scale=0.8]{} ++(\pagesize*4.5-0.5,1.2);
    \draw[rounded corners=0pt, dotted] (0+\pagesize*0,-2.6) rectangle node[scale=0.5]{} ++(\pagesize*4.5,0.4);

    \draw[rounded corners=0pt, fill=fancygray!15] (0+\pagesize*2,-0.8) rectangle node[scale=0.8]{} ++(\pagesize,1.2);

    \draw[rounded corners=0pt, draw=none] (0+\pagesize*0+0.1,-0.8) rectangle node[scale=0.5, text width=2.7cm, align=center]{Computer's Physical Memory} ++(\pagesize*2,1.2);
    \draw[rounded corners=0pt, draw=none] (0+\pagesize*3,-0.8) rectangle node[scale=0.5, text width=3.5cm, align=center]{Number inside Page represents KeyID} ++(\pagesize*1.3,1.2);

    \draw[rounded corners=0pt, draw=none] (0+\pagesize*0+0.3,-2.6) rectangle node[scale=0.5, text width=2.7cm, align=center]{Virtual Memory} ++(\pagesize,0.4);
    \draw[rounded corners=3pt, draw=none] (0+\pagesize*0.25,\ptrY) rectangle node[scale=0.5, text width=4.3cm, align=center]{Page Table Entries} ++(\pagesize,0.8);

    \draw[rounded corners=4pt, draw=none] (0+\pagesize*2,0.55) rectangle node[scale=0.6]{\SI{4}{\kilo\byte} Page} ++(\pagesize,0);

    \draw[rounded corners=3pt, fill=black!80] (\startg,\ptrY) rectangle node[scale=0.7]{} ++(\pagesize,0.7);
    \draw[rounded corners=3pt, draw=none] (\startg,\ptrY) rectangle node[scale=0.4,text=white]{R/W} ++(\pagesize,0.2);
    \draw[rounded corners=3pt, fill=fancyblue!55] (\startg,-1.8) rectangle node[scale=0.5, text width=1.4cm, align=center]{KeyID\,2} ++(0.75,0.5);
    \draw[rounded corners=3pt, fill=fancygray!15] (\startg+0.75,-1.8) rectangle node[scale=0.6, text width=1cm, align=center]{PPN} ++(0.75,0.5);

    \draw[rounded corners=0pt, fill=fancyblue!55] (\startg,-2.6) rectangle node[scale=0.8]{} ++(\pagesize,0.4);
    \draw[rounded corners=0pt, fill=fancyblue!55] (\startg,-2.6) rectangle node[scale=0.8]{y} ++(\pagesize/2,0.4);
    \draw[rounded corners=0pt, fill=fancyblue!40] (\startg + \pagesize/2,-2.6) rectangle node[scale=0.8](g){} ++(\pagesize/4,0.4);
    \draw[rounded corners=0pt, pattern=north east lines, pattern color=fancyblue!60] (\startg + \pagesize/2,-2.6) rectangle node[scale=0.8](g){} ++(\pagesize/4,0.4);
    \draw[rounded corners=3pt, fill=black!80] (\starth,\ptrY) rectangle node[scale=0.7]{} ++(\pagesize,0.7);
    \draw[rounded corners=3pt, draw=none] (\starth,\ptrY) rectangle node[scale=0.4,text=white]{R/W} ++(\pagesize,0.2);
    \draw[rounded corners=3pt, fill=shared] (\starth,-1.8) rectangle node[scale=0.5, text width=1.4cm, align=center]{KeyID\,3} ++(0.75,0.5);
    \draw[rounded corners=3pt, fill=fancygray!15] (\starth+0.75,-1.8) rectangle node[scale=0.6, text width=1cm, align=center]{PPN} ++(0.75,0.5);

    \draw[rounded corners=0pt, fill=shared!70] (\starth,-2.6) rectangle node[scale=0.8]{} ++(\pagesize,0.4);
    \draw[rounded corners=0pt, pattern=north east lines, pattern color=shared] (\starth,-2.6) rectangle node[scale=0.8]{} ++(\pagesize,0.4);
    \draw[rounded corners=0pt, fill=shared] (\starth + \pagesize/2,-2.6) rectangle node[scale=0.8](i){z} ++(\pagesize/4,0.4);

    \draw[rounded corners=0pt, fill=black!40] (0+\pagesize*2+\pagesize/2,-0.8) rectangle node[scale=0.8]{3} ++(0.1875*2,1.2);
    \draw[rounded corners=0pt, fill=fancygray!15] (0+\pagesize*2+\pagesize*3/4,-0.8) rectangle node[scale=0.8]{2} ++(0.1875*2,1.2);
    \draw[rounded corners=0pt, fill=fancygray!15] (0+\pagesize*2,-0.8) rectangle node[scale=0.8]{2} ++(0.1875*2*2,1.2);

    \draw[edge] (\startg,-2.9) -- ++(0,0.3);
    \draw[rounded corners=4pt, draw=none] (\startg,-3.1) rectangle node[scale=0.6]{{\color{fancyblue}ptrY}} ++(0,0); %

    \draw[edge, fancyred] (\startg+\pagesize/2+0.1,-2.9) -- ++(0,0.3) node[midway,yshift=-0.3em]{\scriptsize\ding{55}};
    \draw[rounded corners=4pt, draw=none] (\startg+\pagesize/2+0.1,-3.1) rectangle node[scale=0.6]{{\color{fancyred}ptrY}} ++(0,0); %

    \draw[edge] (\starth+\pagesize/2,-2.9) -- ++(0,0.3);
    \draw[rounded corners=4pt, draw=none] (\starth+\pagesize/2,-3.1) rectangle node[scale=0.6]{{\color{shareddark}ptrZ}} ++(0,0); %

    \draw[edge] (\startg+\pagesize/2,-1.3) -- ++(0,0.25) -| (\starti+\pagesize/2,-0.8);
    \draw[edge] (\starth+\pagesize/2,-1.3) -- ++(0,0.25) -| (\starti+\pagesize/2,-0.8);

\end{tikzpicture}%
    \caption{
      Overview of \tmebox's sub-page granular isolation.
      Aliasing allows the fine-grained encryption of the sandbox's memory, which is applicable to individual objects located on the same page.
      Here, two sandboxes share the same physical page to store their private data.
      Each sandbox can access memory only with its assigned \keyid through the correct alias.
      Hence, parts of the physical page are encrypted differently, with the respective sandbox's encryption key.
      This limits access for a sandbox to their respective part of the physical page since accessing non-owned data (with the wrong \keyid) is detected by the hardware-based integrity checks.
      }
    \label{fig:tmebox:implementation:design_sub}
  \end{center}
\end{figure}
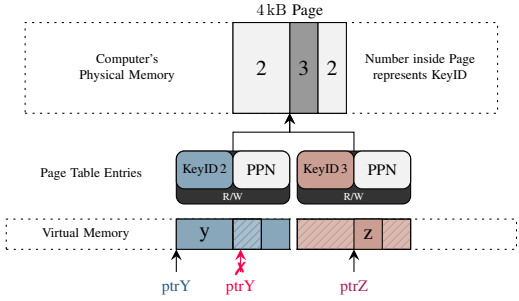

In addition to page granular isolation, our design also enables more fine-grained access control on the level of individual cache lines within a single page.
This sub-page granular isolation is achieved by page aliasing, where two (or more) sandboxes, with different \keyids each, have memory assigned on the same physical page.

\Cref{fig:tmebox:implementation:design_sub} illustrates the sub-page granular isolation of \tmebox, allowing fine-grained resource management for distinct objects associated with a sandbox.
In the depicted example, two sandboxes, \texttt{y} and \texttt{z}, use parts of the memory of the same physical page.
The sandboxes use different \keyids and, thus, their stored data is encrypted using different \tmemk encryption keys.
This example shows that even when multiple sandboxes store their own respective data co-located on a physical page, the data is still isolated from each other.
\tmebox's fine-grained isolation can be applied to individual cache lines, as this is the smallest granularity of \tmemk with integrity.
This means we can isolate memory without constraining memory allocation of the process, \eg by relying on a page granularity.

\begin{figure}
  \begin{center}
    \input{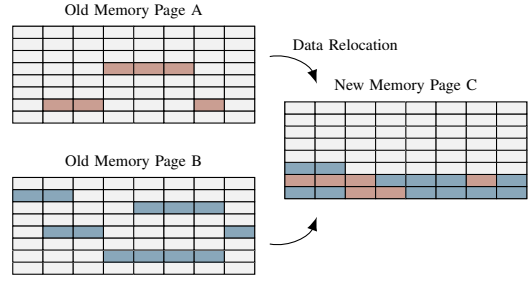}%
    \caption{
      Overview of \tmebox's flexible data relocation.
      This allows for efficient memory migration of sparsely allocated cache lines to different memory locations.
      When reallocating memory objects, the allocator can move allocations from different sandboxes to the same physical page.
      \Eg when a page only contains a single object, an allocator can decide to move the object to another already-used page, such that the old page becomes entirely unused. Then, the now-unused old physical page can be freed and reclaimed for other tasks. %
      } %
    \label{fig:tmebox:implementation:relocation}
  \end{center}
\end{figure}

\paragraph{Flexible Relocation of Data}
Our scheme enables the flexible relocation of data located in memory, which is important for software systems that focus on data-centric computation.
This allows the memory allocator to efficiently migrate memory, thereby reducing the overall memory fragmentation and consumption. %
\Cref{fig:tmebox:implementation:relocation} provides an overview of \tmebox data relocation.
The figure shows the relocation of sparsely allocated cache lines from different sandboxes on memory pages \texttt{A} and \texttt{B} to a single page, page \texttt{C}.
This allows the usage of allocator caches and results in a more efficient management of fragmented memory resources.

In contrast, the partitioning of the virtual address space only allows to operate on contiguous memory regions, \ie on the granularity of a large number of pages.
Such systems cannot migrate allocations from different sandboxes to the same page.
Here, \tmebox's approach offers an advantage by allowing fine-grained and flexible relocation of data across different sandboxes. %

\subsection{Isolation of In-Process Sandboxes}

\tmebox isolates code and data for sandboxing.
This includes the fine-grained isolation of data in memory, restricting access for mutually untrusted sandboxes.
Additionally, \tmebox ensures the isolated execution of sandboxed code.

\paragraph{Integrity Enforcement}
\tmebox repurposes Intel \tmemk to achieve scalable memory isolation.
Any violation of \tmebox's memory access policies, \ie one sandbox trying to access another sandbox's memory, is detected by the \tmemk encryption engine.
\tmebox relies on \tmemk's integrity checks to detect sandbox access violations.
In particular, \tmemk associates a cryptographic MAC that is 28-bit in size with each \SI{64}{\byte} cache line~\cite{inteltdxxwhitepaper}. %

Memory is initialized by writing the entire cache line, which sets up the MAC with the used \keyid, thus providing data integrity.
For subsequent memory reads, \tmemk authenticates the MAC using the stored DRAM data and \keyid of the access.
\tmebox leverages this authentication procedure to detect unauthorized memory accesses of sandboxes.

Memory requests with an incorrect \keyid result in a MAC mismatch, triggering an exception. %
Similarly, (partial) writes to a cache line with an incorrect \keyid corrupts the MAC associated with the cache line.
In this case, the access violation is detected when the sandbox owning the memory location performs a read access.
This allows \tmebox to detect access violations within the security bounds provided by the cryptographic MAC.

\paragraph{Isolated Execution}
In addition to cryptographically isolating memory resources, software sandboxing requires isolating the executed code, \ie control-flow transfers remain within a sandbox's isolated code region. %
\tmebox achieves this through compiler instrumentation, which modifies forward-edge and backward-edge control-flow transfers to enforce this property.

For forward-edge transfers, \tmebox instruments indirect function calls and jumps (\ie function pointer dereferences) to stay within the sandbox's isolated code region.
Address masking is applied to these function pointers, enforcing that code execution is confined within the sandbox.

Similarly, backward-edge transfers, such as function returns, are transformed and instrumented to ensure that return addresses target code within the sandbox.
Additionally, direct function calls are analyzed during compile-time to ensure that they target valid call sites, \ie authorized call targets within the sandbox or trusted runtime calls.

Alternatively, Intel CET~\cite{DBLP:conf/isca/ShanbhogueGS19} can be used to protect control-flow transfers in the \tmebox sandbox.
The CET shadow stack feature can replace software-based return address instrumentation for backward-edge CFI, thus enhancing security and performance by protecting return addresses in hardware.
In terms of forward-edge CFI, the IBT feature of CET would also provide an additional layer of security.
Although indirect function calls would still need to be instrumented to reside within the sandbox code region, the jump targets would be limited to function entries, thus increasing security.

\section{Prototype Implementation}\label{sec:tmebox:implementation}

In this section, we detail our prototype implementation.
The \tmebox framework comprises an LLVM~\cite{DBLP:conf/cgo/LattnerA04} compiler extension and a security-hardened memory allocator.
In addition, we use a Linux kernel patch that enables us to control the \tmemk encryption engine.

\subsection{Compiler Extension}
We base our prototype implementation on the LLVM compiler infrastructure~\cite{DBLP:conf/cgo/LattnerA04} (version 14.0.0).
Our compiler extension consists of a set of individual compiler passes required for \tmebox's isolation of code and data.
Moreover, the \tmebox framework is fully parameterizable regarding the utilized virtual address space (\eg 57-bit virtual addressing).

We also detail architecture-specific optimizations in our prototype implementation, resulting in more efficient code instrumentation. %
Specifically, \tmebox integrates the following three compiler modifications:

\paragraph{CPU Register}
First, we need to reserve a dedicated CPU register to securely store the base address of the sandbox and enable fast access.
\tmebox implements two options for this:
We either apply an x86-specific optimization by using a segment register\footnote{The x86 architecture supports segmentation, allowing memory instructions to use a segment-based addressing mode. In particular, the \texttt{fs} and \texttt{gs} segment registers are still functional in 64-bit mode. While the \texttt{fs} segment is commonly used to address thread local storage (TLS), the \texttt{gs} segment has no common use and, thus, can be utilized by applications~\cite{segementregister}.}~\cite{segementregister} with segment-relative memory accesses or reserve a general-purpose register.
Our compiler framework supports both options, allowing for maximal compatibility and performance comparison, which is reflected in our evaluation.

Depending on the chosen \tmebox mode, we either reserve the \texttt{gs} segment register or a single general-purpose register, in our case \texttt{r15}, to store the sandbox's base address (\ie the upper part of the virtual address) that corresponds to the sandbox's assigned \keyid in the PTEs, which maps to the designated encryption key of the sandbox.
Note that reserving a general-purpose CPU register has performance implications due to increased register pressure.
Future \tmebox implementations can use the Intel advanced performance extensions~(APX)\footnote{Intel Advanced Performance Extensions~(APX)~\cite{intelapx} enhance x86 processors by increasing the number of general-purpose registers from 16 to 32.}~\cite{intelapx} that enhance x86 processors with an increased register set.
This optimization would then also lessen the overhead of reserving a general-purpose register.

\paragraph{Memory Operations}
Next, we integrate a compiler pass that transforms memory operations, enabling the control of the base address and index of memory accesses. %
The compiler instruments every memory operation to enforce that accesses use the sandbox's designated encryption key (through the sandbox's \keyid).
For instructions that operate with memory (\eg loads and stores), the corresponding memory operands are truncated, and the sandbox's base address is instrumented.
\tmebox is parameterizable, depending on the size of the virtual address space and the machine's available \tmemk keys.
Depending on the number of keys, the pointer is truncated by clearing the uppermost canonical bits of the address, starting with the highest order bit available. %
Subsequently, the sandbox's base address, stored in the reserved register (\ie either \texttt{gs} or \texttt{r15}), is instrumented.
This is achieved through segment-based addressing or instruction insertion.

Our prototype uses 57-bit virtual addressing, and the Intel Xeon Gold 6530 processor used in our evaluation supports 6-bit \keyids.
In this configuration, we clear the topmost 16 bits of a 64-bit pointer, resulting in 48-bit addresses within sandboxes, leaving enough canonical bits to place the base address in the virtual address pointer. %
Note that \tmemk supports up to 15-bit \keyids, and the number of implemented encryption keys is platform-dependent.
Since the CPU used in our evaluation supports 6-bit \keyids, we can operate 63 sandboxes (excluding the default \keyid 0). %
However, for future hardware that supports more \keyids, our compiler framework can be reconfigured to truncate up to 15 bits of the canonical address to encode the sandbox base address.
This truncation is performed by applying a mask to the pointer.

Special care is required for the \texttt{eflags}\footnote{The \texttt{eflags} register of the x86 architecture contains the current CPU state represented by status flags, control flags, and system flags~\cite{intelguide-2a,intelguide-3a}.} register of the x86 architecture. %
The compiler pass checks whether the targeted instructions use \texttt{eflags} before instrumenting an instruction sequence.
If so, we rely on an instruction sequence that does not alter the \texttt{eflags}, as saving and restoring them would induce non-negligible performance overheads.
For the \texttt{gs} mode, LLVM provides an interface to transform memory operations to segment-based addressing.
In \texttt{r15} mode, the base address is inserted with an additional instruction.
We also implement a common optimization for stack pointer-relative accesses similar to existing work~\cite{DBLP:conf/sp/YeeSDCMOONF09, DBLP:conf/uss/SehrMBKPSYC10}.
To use this optimization, we need to instrument the register \texttt{rsp} and \texttt{rbp} whenever they are restored from the stack, as they could be tampered with.
Besides stack-relative accesses, all other memory operations are instrumented.

\paragraph{Control-Flow Transfers}
Lastly, we implement a compiler pass that isolates the code of the sandbox by instrumenting control-flow transfers.
Precisely, forward-edge and backward-edge transitions are instrumented via address masking, restricting transfers to remain in the sandbox's code region, comparable to recent SFI-based sandboxing~\cite{DBLP:conf/sp/YeeSDCMOONF09, DBLP:conf/uss/SehrMBKPSYC10}.

For forward-edge transfers (\ie indirect function calls and jumps), the compiler instruments the corresponding function pointers before executing the dedicated instruction. %
This restricts addressable call sites to the virtual address space of the respective sandbox.
Similarly, backward-edge transfers (\ie function returns) are instrumented.
Here, our compiler transforms returns into a code sequence that receives the return address from the stack, instruments it, and performs an indirect jump.
As an optimization, a future \tmebox implementation can use Intel~CET\footnote{Intel Control-flow Enforcement Technology~(CET)~\cite{DBLP:conf/isca/ShanbhogueGS19} integrates a shadow stack feature that applies return address protection in hardware.}~\cite{DBLP:conf/isca/ShanbhogueGS19} for return address protection, increasing security and enhancing system performance.
Additionally, direct function calls are checked on compile-time for valid call sites, \ie the sandbox is allowed to call this function directly.

\subsection{Memory Allocator}
The memory management of \tmebox is responsible for initializing memory locations of runtime data (\eg stack and heap memory) of individual sandboxes.
This initialization is achieved by writing the entire \SI{64}{\byte} cache line with the associated \keyid.
For stack and static memory, this is done during the start-up procedure of the sandbox.
The sandbox's stack memory region is mapped and initialized with the corresponding \keyid of the sandbox.
Moreover, static memory is copied into the sandbox's dedicated virtual memory region, \ie the region is mapped and the corresponding \keyid is assigned.
Also, the sandbox base address is set up in the dedicated register, \ie either \texttt{gs} or \texttt{r15}, of the sandbox.

Furthermore, the memory allocator is responsible for initializing and managing heap memory. %
Therefore, the memory allocator maps the heap for every sandbox (with their designated \keyid) and aliases it to the same physical memory region.
Moreover, the memory allocator then performs the initialization of the corresponding heap memory on allocation.
This procedure enables the memory allocator to apply the scalable isolation and flexible relocation of data in memory.
Precisely, the memory allocator can use fine-grained isolation, ranging from individual cache lines to a large number of pages, depending on the runtime constraints.
On pages that contain data from multiple sandboxes, the allocator needs to align memory chunks to cache line size.
Here, our design enables the relocation of data, as individual memory chunks can be immediately reallocated.
Note that data from different sandboxes is never co-located within a single cache line.

\subsection{Linux Kernel Patch}

The operating system needs to provide software support to control the \tmemk hardware feature.
For our implementation, we use an experimental patch for the Linux kernel provided by Intel Labs that enables the setting of \keyids in the PTEs~\cite{tmemkpatch}.
This kernel patch provides a syscall interface that allows additional arguments for the \texttt{mprotect} system call to associate a specific \keyid with a page (akin to Intel MPK's \texttt{pkey\_mprotect}).

\begin{figure}
  \begin{center}
    \definecolor{fancyred}{HTML}{f70146}
\definecolor{fancypurple}{HTML}{6c2f91}
\definecolor{fancymid}{HTML}{5191c1}
\definecolor{fancygray}{HTML}{a5a5a5}
\definecolor{fancyblue}{HTML}{285f82}
\definecolor{fancygreen}{HTML}{78b473}
\definecolor{fancyyellow}{HTML}{ff8c00}
\definecolor{fancycyan}{HTML}{77babf}
\definecolor{fancyhead}{HTML}{245b78}
\definecolor{fancybody}{HTML}{e2e9ed}

\begin{tikzpicture}
    \tikzset{edge/.style=->, >=stealth, thick};

    \draw[rounded corners=0pt, draw=none, fill=fancygray!30] (0,-0.6) rectangle node[scale=0.7]{Physical Page Number} ++(8,0.6);
    \draw[rounded corners=0pt, draw=none, fill=fancygray!30] (7.5,0) rectangle node[scale=0.6]{} ++(0.5,0.3);
    \draw[rounded corners=0pt, draw=none, fill=fancygray!30] (0,-0.9) rectangle node[scale=0.6]{} ++(2,0.3);
    \draw[rounded corners=0pt] (0,0.3) rectangle node[scale=0.6]{} ++(8,-1.2);

    \draw[rounded corners=0pt, fill=fancygray!15] (0,0) rectangle node[scale=0.7]{XD} ++(0.5,0.3);
    \draw[rounded corners=0pt, fill=fancygray!15] (0.5,0) rectangle node[scale=0.7]{Protection Key} ++(2.0,0.3);
    \draw[rounded corners=0pt, fill=fancygray!15] (2.5,0) rectangle node[scale=0.7]{Ignored} ++(3.5,0.3);
    \draw[rounded corners=0pt, fill=fancygray!15] (6,0) rectangle node[scale=0.6]{RSVD} ++(1.0,0.3);
    \draw[rounded corners=0pt, fill=fancyyellow!20] (7,0) rectangle node[scale=0.7]{} ++(1,0.3);
    \draw[rounded corners=0pt, fill=fancyyellow!20] (0,-0.3) rectangle node[scale=0.7]{KeyID} ++(2.0,0.3);

    \draw[rounded corners=0pt, fill=fancygray!15] (2.0,-0.9) rectangle node[scale=0.7]{R} ++(0.5,0.3);
    \draw[rounded corners=0pt, fill=fancygray!15] (2.5,-0.9) rectangle node[scale=0.7]{Ignored} ++(1.0,0.3);
    \draw[rounded corners=0pt, fill=fancygray!15] (3.5,-0.9) rectangle node[scale=0.7]{G} ++(0.5,0.3);
    \draw[rounded corners=0pt, fill=fancygray!15] (4.0,-0.9) rectangle node[scale=0.6]{PAT} ++(0.5,0.3);
    \draw[rounded corners=0pt, fill=fancygray!15] (4.5,-0.9) rectangle node[scale=0.7]{D} ++(0.5,0.3);
    \draw[rounded corners=0pt, fill=fancygray!15] (5.0,-0.9) rectangle node[scale=0.7]{A} ++(0.5,0.3);
    \draw[rounded corners=0pt, fill=fancygray!15] (5.5,-0.9) rectangle node[scale=0.6]{PCD} ++(0.5,0.3);
    \draw[rounded corners=0pt, fill=fancygray!15] (6.0,-0.9) rectangle node[scale=0.6]{PWT} ++(0.5,0.3);
    \draw[rounded corners=0pt, fill=fancygray!15] (6.5,-0.9) rectangle node[scale=0.7]{US} ++(0.5,0.3);
    \draw[rounded corners=0pt, fill=fancygray!15] (7.0,-0.9) rectangle node[scale=0.7]{RW} ++(0.5,0.3);
    \draw[rounded corners=0pt, fill=fancygray!15] (7.5,-0.9) rectangle node[scale=0.7]{P} ++(0.5,0.3);

    \draw[rounded corners=4pt, draw=none] (0.05,0.45) rectangle node[scale=0.6]{63} ++(0,0);
    \draw[rounded corners=4pt, draw=none] (7.95,0.45) rectangle node[scale=0.6]{48} ++(0,0);
    \draw[rounded corners=4pt, draw=none] (7.0,0.45) rectangle node[scale=0.6]{Max\_PA} ++(0,0);
    \draw[rounded corners=4pt, draw=none] (0.05,-1.05) rectangle node[scale=0.6]{15} ++(0,0);
    \draw[rounded corners=4pt, draw=none] (7.95,-1.05) rectangle node[scale=0.6]{0} ++(0,0);
\end{tikzpicture}%
    \caption{The page table entries on the Intel x86-64 architecture~\cite{intelguide-3a}. Intel \tmemk applies changes to the specification of physical address~\cite{intelmktme}. In particular, \tmemk repurposes the upper bits of the physical address to carry the \keyid to the encryption engine in the memory controller, resulting in a reduction of addressable physical memory.} %
    \label{fig:tmebox:implementation:pte}
  \end{center}
\end{figure}
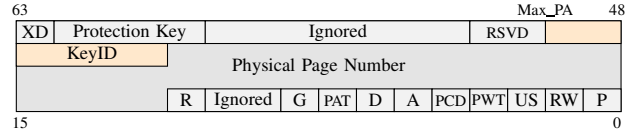

\Cref{fig:tmebox:implementation:pte} illustrates the PTEs on the Intel x86-64 architecture~\cite{intelguide-3a}, including the changes in the specification to support \tmemk~\cite{intelmktme}.
In particular, \tmemk repurposes the upper bits of the physical address, starting with the highest order bit available, to encode the \keyid into the PTE.
Thereby, the \keyid can be transferred to the encryption engine in the memory controller, performing the encryption procedure.
Note that this reduces the addressable physical memory by the number of \keyid bits in use. %

\section{Security Analysis}\label{sec:tmebox:security_analysis}

In this section, we provide an in-depth security analysis of our \tmebox design.
We comprehensively analyze potential security threats and detail the derived security properties of the \tmebox sandbox against an attacker, defined in our thread model (\cf \Cref{sec:tmebox:threat_model}).

\subsection{Systematic Analysis}

The generic attack path consists of one or several memory safety vulnerabilities present in the target unprivileged user space application.
We assume that the exploitable memory safety error grants the attacker arbitrary read and write capabilities.
The attacker can then use this primitive in an attempt to leak or corrupt data in memory, exploiting the software system.

\paragraph{In-Process Memory Isolation}
We use \tmebox to isolate individual software components of the application to minimize the attack surface of software exploitation.
Consequently, the memory safety error cannot be exploited to escape the isolated sandbox, \ie the attacker cannot leak or corrupt the memory of other sandboxes.

We achieve this by leveraging compiler instrumentation.
\tmebox ensures that the sandbox uses its designated encryption key for memory operations.
By controlling the base and index of memory operands, accesses use the sandbox's \keyid, which corresponds to the sandbox's encryption key.
Note that no memory operation exists within a sandbox that bypasses this instrumentation, as this would bypass the protection of the sandbox (\ie a sandbox escape).
Furthermore, we detect memory accesses outside the sandbox's assigned memory with \tmemk's cryptographic integrity checks since the sandbox is forced to access memory with its encryption key.
Specifically, the \tmemk encryption engine verifies the sandbox's memory access.
Thus, a violation is detected by the MAC authentication procedure.

Moreover, \tmebox allows for flexible memory management.
When a sandbox allocates memory, it is initialized with the sandbox encryption key through a memory write with the corresponding \keyid.
Subsequently, access to this memory location is solely granted with that specific encryption key (through the \keyid) associated with that sandbox.
\tmebox supports isolation granularities ranging from single cache lines to full pages.
The smallest granularity supported by our isolation mechanism is a \SI{64}{\byte} cache line, as this is the granularity of the TME-MK integrity checks.
Similarly, when memory has been freed by one sandbox and is allocated by another, the memory location is (re-) initialized with the \keyid of the new sandbox.
Consequently, this memory location can then be accessed exclusively by the new sandbox.

\paragraph{Cryptographic Integrity}
The \tmemk encryption engine provides integrity protection for DRAM memory.
Once a memory location (\ie cache line) is initialized with a specific \keyid, it is protected by the MAC, which preserves integrity at the cache line level. %
During memory reads, the memory controller requests the DRAM data and its associated MAC, verifying the memory access.

Successfully verified memory accesses can result in the caching of the data, with the corresponding cache line marked with the \keyid used for the access.
Any attempt to access memory with an incorrect \keyid is detected by the MAC verification within cryptographic bounds.
Specifically, cache lines are tagged with the physical address, and the \keyid is a part of that.
Thus, accessing cache lines with a different \keyid results in a cache miss and leads to a DRAM access.
This DRAM access is then authenticated by the encryption engine, which detects the usage of incorrect \keyids.

Writing to memory with an incorrect \keyid corrupts the associated MAC.
Although this will not immediately result in the detection of the violation, this corruption is also not security critical since no secret information can be extracted.
Subsequent memory reads from this corrupted memory location will trigger a MAC authentication, thus detecting the corruption of the sandbox's memory.
In such cases, the trusted runtime handles the exception, preventing the attacker from exploiting the system.

\tmebox leverages Intel \tmemk's integrity enforcement, which is based on the security of a cryptographic MAC.
Particularly, \tmemk uses a 28-bit MAC generated with the cryptographically secure hash algorithm SHA-3~\cite{guido2011k}.
The MAC verification ensures that any memory access outside the sandbox's assigned memory is detected with the probability of $1 - 2^{-28}$, resulting in an exception.
Note that attackers cannot bypass this cryptographic integrity check, as our scheme enforces the use of the sandbox's corresponding \keyid for every memory operation, preventing the exploitation of maliciously crafted pointers.
Although a MAC collision could theoretically occur with a probability of $2^{-28}$, the attacker would only receive wrongly decrypted, mangled data, thereby preserving the confidentiality of the data in memory.
Hence, an attacker would not obtain meaningful information on the correct unencrypted data. %

\paragraph{Control-Flow Integrity}
Software sandboxing techniques need to ensure that the executed program remains in its designated code region.
Thus, control-flow transfers must be restricted so that indirect function calls and function returns are limited to code within the sandbox.
\tmebox enforces these restrictions by limiting control-flow transfers to the sandbox code region, similar to recent software-based fault isolation~(SFI) techniques~\cite{DBLP:journals/ftsec/Tan17}.
Indirect function calls and function returns are instrumented, applying address masking to enforce that the call site is within the sandbox code region.
Additionally, direct function calls are analyzed during compile time to verify that all direct call targets are valid.

Moreover, software sandboxing typically also ensures that untrusted sandboxed code within the process has only restricted access to system calls.
These system calls are then executed by a trusted runtime, which can be invoked by the sandbox through a defined interface.
This includes critical system calls, such as \texttt{mprotect} and \texttt{mmap}, used for managing memory resources.
Another attack path for memory corruption consists of gaining an arbitrary read and write primitive followed by remote code execution~(RCE), \eg by invoking the \texttt{exec} system call.
This is followed by a privilege escalation attack to compromise the entire software system~\cite{DBLP:conf/ccs/ZengLLX0DSB23,DBLP:conf/uss/MaarGUOM24,DBLP:conf/ccs/LinWX22}.
To mitigate this security threat, the usage of system calls is only allowed for the trusted runtime, \ie the sandbox has to invoke system calls through the trusted runtime. %
Also, the no-execution policy on writable memory (\ie Write-XOR-Execute) prevents code injection attacks~\cite{DBLP:journals/ieeesp/Arce04b}.
In addition, orthogonal measures such as syscall filtering~\cite{DBLP:conf/uss/SchrammelWSM22} can be employed to further restrict access to system calls.

\paragraph{Multithreading}
Security measures must be designed to operate effectively with multithreaded software systems.
Ensuring thread safety is crucial to prevent concurrency attacks, exploiting time-of-check to time-of-use~(TOCTTOU)~\cite{DBLP:conf/fast/WeiP05} vulnerabilities, where an attacker can misuse a time window between security checks to bypass security measures.

\tmebox addresses this security threat since the \tmemk encryption engine verifies each memory access for read operations in hardware.
Moreover, \tmemk updates the MAC for cache lines whenever data is written to memory.
Specifically, when the CPU requests memory from DRAM, the memory controller fetches the corresponding data and its associated MAC.
\tmemk then verifies whether the memory access uses the correct \keyid and subsequently caches the data if the MAC verification is successful.
Cache lines are tagged with their respective \keyid, enabling access to cached data using the appropriate \keyid.

When performing data relocation, the memory location is initialized with a new \keyid, forcing a writeback for the data in memory that also updates the MAC.
This procedure  ensures that access for sandboxes with revoked \keyids becomes invalid. %
As a result, any subsequent attempts to access memory locations with the revoked \keyid trigger MAC verification errors. %
Note that the processor ensures cache and TLB coherency, maintaining consistency for the cached data and \keyids located in the PTEs across all CPU cores.

\section{Evaluation}\label{sec:tmebox:evaluation}

This section discusses the evaluation of \tmebox in terms of system performance.
Moreover, we evaluate the memory latency of the \tmemk hardware feature.

\paragraph{Evaluation Setup}
All evaluations are performed on an off-the-shelf Intel Xeon Gold 6530 processor, as this model supports Intel \tmemk.
Our system features the following specifications: each of the 32 cores is equipped with a \SI{48}{\kilo\byte} L1D cache, a \SI{32}{\kilo\byte} L1I cache, and a \SI{2}{\mega\byte} L2 cache.
The cores share a \SI{160}{\mega\byte} L3 cache, serving as last-level cache~(LLC).
Furthermore, our system uses \SI{512}{\giga\byte} DDR5-4800 memory with ECC.
Unless stated otherwise, the system is configured to enable Intel \tmemk with integrity support. %

\subsection{Performance Evaluation}
In this section, we conduct the performance evaluation of our \tmebox design.
For our evaluation, we use the SPEC CPU2017~\cite{DBLP:conf/wosp/BucekLK18} benchmark suite and NGINX. %
All benchmarks are compiled with the \texttt{-O3} optimization level of \texttt{clang}.

\paragraph{SPEC CPU2017 Results}
We benchmark one baseline configuration and two \tmebox configurations to showcase the performance overheads.
Note that we use the \emph{ref} input of SPEC CPU2017 for all benchmarks.
Specifically, we evaluate both configurations of \tmebox: \tmebox~(\texttt{gs}) with segment-based addressing and \tmebox~(\texttt{r15}) reserving a general-purpose register, as detailed in the implementation section (\cf \Cref{sec:tmebox:implementation}).
In addition, for both configurations, we further distinguish between data isolation, and code and data isolation.
Our compiler toolchain targets the hardening of C and C++ applications; thus, we exclude all Fortran benchmarks.

\begin{figure}
  \begin{center}
    \begin{tikzpicture}

    \begin{axis} [
        ybar = 0pt,
        height=3.8cm,
        width=1.0\columnwidth,
        bar width = 4pt,
        ylabel = {Performance Overhead [\%]},
        ylabel style={yshift=-1.5em},
        ymin = 0,
        ymax = 40,
        ytick={0,10,20,30,40},
        ymajorgrids,
        yminorgrids,
        minor grid style={dashed,black!30},
        major grid style={dashed,black!30},
        enlarge x limits={0.035},
        xtick=data,
        symbolic x coords = {600.perlbench\_s,602.gcc\_s,605.mcf\_s,619.lbm\_s,631.deepsjeng\_s,638.imagick\_s,644.nab\_s,657.xz\_s,500.perlbench\_r,502.gcc\_r,505.mcf\_r,519.lbm\_r,531.deepsjeng\_r,538.imagick\_r,544.nab\_r,557.xz\_r,Geomean SPEC},
        y label style={font=\scriptsize},
        x tick label style={rotate=30,anchor=east,font=\scriptsize},
        y tick label style={font=\scriptsize},
        visualization depends on={value \thisrow{text shift} \as \ErrorShift},
        every node near coord/.append style={/pgf/number format/1000 sep=, rotate=90, anchor=west, xshift=\ErrorShift, font=\scriptsize,/pgf/number format/.cd,fixed,fixed,fixed zerofill,precision=1},
        legend columns=4,
        legend style={
          at={(1,1.3)},
          legend cell align=left,
          font=\tiny,
        },
        color=black,
    ]
    \addplot+ [
      black,fill=black!5,%
    ] table [col sep=semicolon,x=name,y=overhead, y error=sterr] {results/config_gs.csv};
    \addlegendentry{\tmebox(\texttt{gs}) \textsubscript{data}};

    \addplot+ [
      black,fill=black!10,postaction={pattern=north east lines, pattern color=black!60},
    ] table [col sep=semicolon,x=name,y=overhead, y error=sterr] {results/config_gs_cfi.csv};
    \addlegendentry{\tmebox(\texttt{gs}) \textsubscript{code+data}};

    \end{axis}

\end{tikzpicture}%
    \caption{The relative performance overhead of \tmebox using the \texttt{gs}-mode for the SPEC CPU2017 benchmark suite.}
    \label{fig:tmebox:evaluation:performance:gs}
  \end{center}
\end{figure}
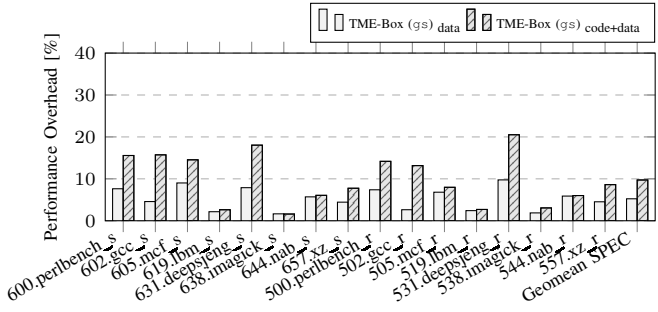

\begin{figure}
  \begin{center}
    \begin{tikzpicture}

    \begin{axis} [
        ybar = 0pt,
        height=3.8cm,
        width=1.0\columnwidth,
        bar width = 4pt,
        ylabel = {Performance Overhead [\%]},
        ylabel style={yshift=-1.5em},
        ymin = 0,
        ymax = 40,
        ytick={0,10,20,30,40},
        ymajorgrids,
        yminorgrids,
        minor grid style={dashed,black!30},
        major grid style={dashed,black!30},
        enlarge x limits={0.035},
        xtick=data,
        symbolic x coords = {600.perlbench\_s,602.gcc\_s,605.mcf\_s,619.lbm\_s,631.deepsjeng\_s,638.imagick\_s,644.nab\_s,657.xz\_s,500.perlbench\_r,502.gcc\_r,505.mcf\_r,519.lbm\_r,531.deepsjeng\_r,538.imagick\_r,544.nab\_r,557.xz\_r,Geomean SPEC},
        y label style={font=\scriptsize},
        x tick label style={rotate=30,anchor=east,font=\scriptsize},
        y tick label style={font=\scriptsize},
        visualization depends on={value \thisrow{text shift} \as \ErrorShift},
        every node near coord/.append style={/pgf/number format/1000 sep=, rotate=90, anchor=west, xshift=\ErrorShift, font=\scriptsize,/pgf/number format/.cd,fixed,fixed,fixed zerofill,precision=1},
        legend columns=4,
        legend style={
          at={(1,1.3)},
          legend cell align=left,
          font=\tiny,
        },
        color=black,
    ]
  	\addplot+ [
      black,fill=black!60,%
  	] table [col sep=semicolon,x=name,y=overhead, y error=sterr] {results/config_r15.csv};
  	\addlegendentry{\tmebox(\texttt{r15}) \textsubscript{data}};

    \addplot+ [
      black,fill=black!70,postaction={pattern=north east lines, pattern color=black!1},
    ] table [col sep=semicolon,x=name,y=overhead, y error=sterr] {results/config_r15_cfi.csv};
    \addlegendentry{\tmebox(\texttt{r15}) \textsubscript{code+data}};

    \end{axis}

\end{tikzpicture}%
    \caption{The relative performance overhead of \tmebox using the \texttt{r15}-mode for the SPEC CPU2017 benchmark suite.}
    \label{fig:tmebox:evaluation:performance:r15}
  \end{center}
\end{figure}
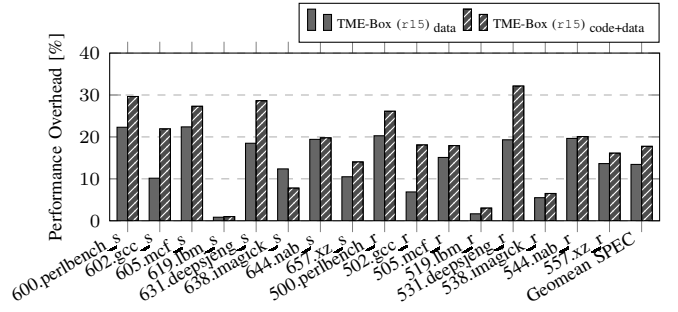

\Cref{fig:tmebox:evaluation:performance:gs} and \Cref{fig:tmebox:evaluation:performance:r15} illustrate the relative performance overhead of \tmebox in \texttt{gs}-mode and \texttt{r15}-mode, respectively, for the SPEC CPU2017 benchmark suite. %
Notably, the \tmebox~(\texttt{gs}) configuration outperforms the \tmebox~(\texttt{r15}) configuration significantly.
We find that \tmebox~(\texttt{gs}) imposes a low geomean overhead of \SPECgs compared to \tmebox~(\texttt{r15}), which imposes a geomean overhead of \SPECr for data isolation.

In addition, \tmebox requires restricting control-flow transfers for its security.
For both code and data isolation, the overhead for \tmebox~(\texttt{gs}) increases to \SPECgscfi, while the overhead of \tmebox~(\texttt{r15}) increases to \SPECrcfi.
Control-flow instrumentation is particularly sensitive to the number of function calls, as these control-flow transfers are instrumented.
We find that the majority of the overhead from code isolation stems from the return address instrumentation.
Specifically, the enabled control flow instrumentation imposes the greatest increase of performance overhead for benchmarks that perform a larger relative number of function calls and returns.

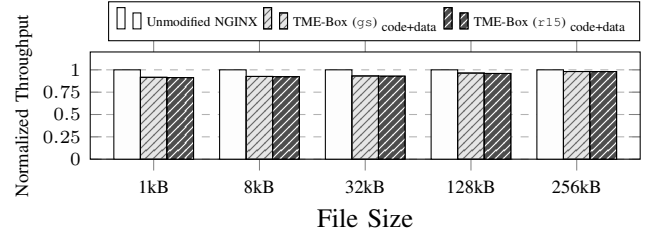
\begin{figure}
  \begin{center}
    \begin{tikzpicture}

    \begin{axis} [
        ybar = 0pt,
        height=3.0cm,
        width=1.0\columnwidth,
        bar width = 10pt,
        ylabel = {Normalized Throughput},
        xlabel={File Size},
        ylabel style={yshift=-1.0em},
        ymin = 0,
        ymax = 1.2,
        ytick={0,0.25,0.5,0.75,1},
        ymajorgrids,
        yminorgrids,
        minor grid style={dashed,black!30},
        major grid style={dashed,black!30},
        enlarge x limits={0.15},
        xtick=data,
        symbolic x coords = {1kB,8kB,32kB,128kB,256kB},
        y label style={font=\scriptsize},
        y tick label style={font=\scriptsize},
        visualization depends on={value \thisrow{text shift} \as \ErrorShift},
        every node near coord/.append style={/pgf/number format/1000 sep=, rotate=90, anchor=west, xshift=\ErrorShift, font=\scriptsize,/pgf/number format/.cd,fixed,fixed,fixed zerofill,precision=1},
        legend columns=-1,
        x tick label style={font=\scriptsize},
        y tick label style={font=\scriptsize},
        legend style={
          at={(1,1.1)},
          legend cell align=left,
          font=\tiny,
          anchor=south east,
        },
    ]
    \addplot+ [
      black,fill=fancyyellow!20,%
      black,fill=black!1,%
    ] table [col sep=semicolon,x=name,y=overhead, y error=sterr] {results/nginx_base.csv};
    \addlegendentry{Unmodified NGINX};

    \addplot+ [
      black,fill=black!10,postaction={pattern=north east lines, pattern color=black!60},
    ] table [col sep=semicolon,x=name,y=overhead, y error=sterr] {results/nginx_gs_cfi.csv};
    \addlegendentry{\tmebox(\texttt{gs}) \textsubscript{code+data}};

    \addplot+ [
      black,fill=black!70,postaction={pattern=north east lines, pattern color=black!1},
    ] table [col sep=semicolon,x=name,y=overhead, y error=sterr] {results/nginx_r15_cfi.csv};
    \addlegendentry{\tmebox(\texttt{r15}) \textsubscript{code+data}};

    \end{axis}

\end{tikzpicture}%
    \caption{Throughput of NGINX with \tmebox for requesting different file sizes normalized to the unmodified NGINX.}
    \label{fig:tmebox:evaluation:performance:nginx}
  \end{center}
\end{figure}

\paragraph{NGINX Results}
Additionally, we perform an evaluation of the NGINX web server (version 1.26.0) using an experimental setup comparable to prior work~\cite{DBLP:conf/ccs/Xie0ZXLKW022,DBLP:conf/uss/Vahldiek-Oberwagner19,DBLP:conf/asplos/NarayanGTRM0FVL23}.
This experiment uses ApacheBench~(\texttt{ab}) to generate requests to receive files of different sizes from the NGINX web server.
We compile NGINX with our \tmebox compiler toolchain and run a single NGINX worker pinned to an isolated CPU core.
We use ApacheBench to perform and benchmark 2,000,000 requests from one client with increasing file sizes covering \SI{1}{\kilo\byte}, \SI{8}{\kilo\byte}, \SI{32}{\kilo\byte}, \SI{128}{\kilo\byte}, and \SI{256}{\kilo\byte}. %
\Cref{fig:tmebox:evaluation:performance:nginx} shows the throughput of NGINX with \tmebox in both modes~(\texttt{gs} and \texttt{r15}) for code and data isolation normalized to the performance of an unmodified NGINX version.
Here, the decrease in throughput for NGINX ranges from \NGINXgsmax to \NGINXgsmin for \tmebox~(\texttt{gs}) and from \NGINXrmax to \NGINXrmin for \tmebox~(\texttt{r15}).
We confirm the observations of prior work~\cite{DBLP:conf/ccs/Xie0ZXLKW022,DBLP:conf/uss/Vahldiek-Oberwagner19} that the measured throughput decreases stronger for smaller files than for larger files, \ie the overhead declines as the file size increases. %

\subsection{Memory Latency}
Besides evaluating the performance overhead, we also measure the additional memory latency imposed by the Intel \tmemk memory encryption with integrity.
We use the \texttt{lat\_mem\_rd} benchmark of the LMBench~\cite{DBLP:conf/usenix/McVoyS96} suite.

\begin{figure}
  \begin{center}
    \definecolor{mathplotlibblue}{HTML}{1f77b4}
\definecolor{mathplotliborange}{HTML}{ff7f0e}

\begin{tikzpicture}

\tikzset{
  testx mark set/.style n args={3}{
    mark=#1,
    draw=#2,
    #3,
    fill=none,
    every mark/.append style={fill=#2,solid,draw=black,line width=0.1pt,scale=0.9}
  }
}

\pgfplotscreateplotcyclelist{mycolorlisttest}{%
  testx mark set={*}{mathplotlibblue}{solid}\\
  testx mark set={square*}{mathplotliborange}{densely dashed}\\
  testx mark set={triangle*}{orange}{dashed}\\
  testx mark set={diamond*}{magenta}{dashdotted}\\
  testx mark set={*}{blue}{dashed}\\
  testx mark set={square*}{red}{dashed}\\
  testx mark set={triangle*}{orange}{dashed}\\
  testx mark set={diamond*}{magenta}{dashed}\\
}

\begin{axis}[
    smooth,
    height=4.2cm,
    width=1.0\columnwidth,
    xlabel={\SI{}{\mega\byte}},
    ylabel={\SI{}{\nano\second}},
    ylabel style={yshift=-1.5em},
    ytick={0,25,50,75,100},
    xmode=log,
    grid=major,
    height=4.2cm,
    width=\hsize,
    legend columns=-1,
    x tick label style={font=\scriptsize},
    y tick label style={font=\scriptsize},
    legend style={
      at={(1,1.1)},
      legend cell align=left,
      font=\tiny,
      anchor=south east,
    },
    cycle list name=mycolorlisttest,
]

\addplot+ [each nth point={10},line width=0.25mm] table [x index=0, y index=1, col sep=space] {results/lmbench_plain.csv};
\addlegendentry{Baseline};

\addplot+ [each nth point={10},line width=0.25mm] table [x index=0, y index=1, col sep=space] {results/lmbench_tmemk.csv};
\addlegendentry{TME-MK};

\draw [color=gray,dashed] (axis cs:0.046875,0) ++(0,-0.2cm) -- ++(0,3cm) node[pos=0.5,anchor=east] {\small L1D};
\draw [color=gray,dashed] (axis cs:2,0) ++(0,-0.2cm) -- ++(0,3cm) node[pos=0.5,anchor=east] {\small L2};
\draw [color=gray,dashed] (axis cs:160,0) ++(0,-0.2cm) -- ++(0,3cm) node[pos=0.5,anchor=east] {\small L3};

\end{axis}

\end{tikzpicture}%
    \caption{The memory latency of the Intel \tmemk memory encryption measured with LMBench.}
    \label{fig:tmebox:evaluation:lmbench}
  \end{center}
\end{figure}
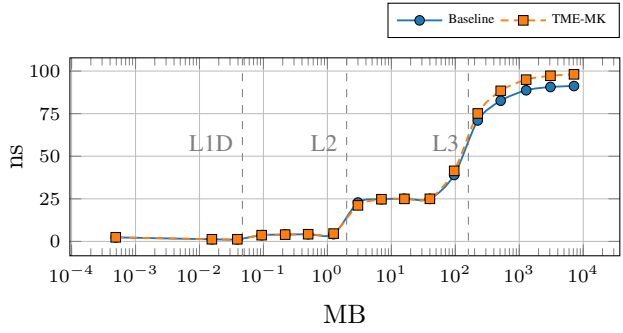

\paragraph{LMBench Results}
We perform the benchmark with a memory size of \SI{8}{\giga\byte} and a stride size of \SI{512}{\byte}.
During benchmarking, the workload is pinned to an isolated CPU core to ensure no other code is executed on that core.
Our workload is executed using two configurations.
For the first configuration, we enable Intel \tmemk during boot, while the second configuration runs with \tmemk disabled.
\Cref{fig:tmebox:evaluation:lmbench} shows the memory latencies for both variants measured with LMBench.
For the L1D, L2, and L3 caches, the access latencies are equivalent.
However, as soon as the memory requests are served from DRAM, we observe different latencies for the two configurations.
The maximum difference in latency that we observe is \SI{6.8}{\nano\second}. %
Note, however, that this overhead is not imposed on each memory access.
It only applies to memory operations that cannot be served directly from the cache and must perform a load from DRAM.

\section{Discussion}\label{sec:tmebox:discussion}

This section compares our design with existing work on software isolation and discusses potential extensions. %

\subsection{Related Work}

In the following, we provide a detailed comparison of \tmebox with prior work, including address space partitioning, memory protection keys, and other isolation mechanisms with page metadata.
Additionally, we compare our work with tag-based isolation schemes and approaches that use cryptographic primitives.

\paragraph{Process Isolation}
Process isolation techniques, such as memory segmentation and paging, separate the memory resources of different processes that are managed by the operating system.
Memory segmentation divides the memory into several regions (\ie segments for code and data), addressing memory locations using a segment identifier and an offset within the segment.
The MMU translates this segment and offset information into a physical address and performs additional access checks (\eg read, write, and execute permissions).
Thus, segmentation allows the operating system to isolate processes, where access to the memory is only granted if the offset is within the segment length and matching access permissions.
Any violation detected by the MMU results in a hardware exception, \ie a segmentation fault is raised.

Furthermore, paging enables process isolation for modern CPUs through the usage of virtual memory.
Paging provides isolation that separates memory access for multiple processes, which is managed on the operating system level.
Specifically, virtual addresses are translated by the MMU, which also checks associated access permissions, thereby preventing illegal access to the memory of other processes.

However, the isolation of individual software components with heavy-weight protection mechanisms, such as process isolation, has an impact on system performance and start-up times~\cite{DBLP:conf/ACMmsp/AikenFHHL06,DBLP:conf/uss/ReisMO19}.
Thus, process isolation is not well suited for the isolation of individual software components of applications. %
As a result, some software systems, like cloud worker architectures~\cite{CFModel}, require more flexible and efficient in-process isolation that allows scalable memory isolation for different sandboxes, as provided by our mechanism.

\paragraph{Software-based Fault Isolation}
Software-based fault isolation~(SFI)~\cite{DBLP:conf/sosp/WahbeLAG93, DBLP:journals/pacmpl/KolosickNJWLGJS22} is a memory isolation technique that divides the virtual address space of a process into predetermined and contiguous regions.
Software instrumentation ensures that memory accesses remain within the isolated data and code regions, thereby enforcing in-process isolation for a sandboxed program.
Traditional SFI relies on address checking or address masking, while modern SFI approaches use guard zones for this data protection~\cite{DBLP:journals/ftsec/Tan17}.

SFI-based isolation has been implemented and evaluated on modern architectures, such as x86 and ARM~\cite{DBLP:conf/sp/YeeSDCMOONF09, DBLP:conf/uss/SehrMBKPSYC10, DBLP:conf/usenix/FordC08,DBLP:conf/sosp/CastroCMPADBB09,DBLP:conf/asplos/Yedidia24,narayan2022segue,DBLP:conf/pldi/MorrisettTTTG12,DBLP:conf/asplos/Yedidia24}.
For instance, Google Native Client~(NaCl)~\cite{DBLP:conf/sp/YeeSDCMOONF09, DBLP:conf/uss/SehrMBKPSYC10} provides sandboxes of \SI{4}{\giga\byte} in size by leveraging software instrumentation and \SI{40}{\giga\byte} guard zones before and after each sandbox.
Moreover, hardware-assisted fault isolation~(HFI)~\cite{DBLP:conf/asplos/NarayanGTRM0FVL23} proposes a processor extension that implements SFI-style isolation efficiently integrated into the processor's architecture, thereby reducing the incurred runtime overhead.

While SFI-style approaches allow for a flexible number of sandboxes, they typically cannot provide fine-grained and dynamic isolation, \ie object-level isolation that can dynamically grow and shrink in size.
In addition, SFI is not designed to isolate non-contiguous memory regions.
Contrarily, \tmebox allows for scalable and fine-grained (\ie sub-page granular) isolation of memory with support for flexible data relocation by repurposing the Intel \tmemk encryption engine.

\paragraph{In-Process Isolation with Page Metadata}
Protection keys for user space~(PKU) is an approach for memory isolation that relies on additional metadata stored in page table entries~(PTE).
The Intel memory protection keys~(MPK)~\cite{DBLP:journals/ieeesp/ParkLK23} allow to dynamically assign a 4-bit protection key to a specific page.
For each memory operation, logical integrity checks are performed that compare the page's protection key against the active access policy located and controlled by the user space protection key register~(PKRU).
The current access permissions (\ie read and write) of each individual protection key can be changed from user space via the \texttt{wrpkru} instruction.
In this way, MPK can be used for data protection, \eg for dynamically locking-away sensitive data in memory like cryptographic key material.

Based on MPK, several academic designs propose software sandboxing~\cite{DBLP:conf/usenix/HedayatiGJCSSM19,DBLP:conf/uss/Vahldiek-Oberwagner19,DBLP:conf/usenix/ParkLXMK19,DBLP:conf/uss/SchrammelWSS0MG20,DBLP:conf/uss/SchrammelWSM22,DBLP:conf/asiaccs/BlairRE23} that facilitates in-process isolation.
However, PKU-based sandboxing can have performance implications in certain scenarios that require frequent sandbox transitions since the \texttt{wrpkru} instruction (including necessary fences) can take more than 100 cycles~\cite{DBLP:conf/uss/Vahldiek-Oberwagner19, DBLP:conf/uss/SchrammelWSS0MG20}.
Similarly, ARMlock~\cite{DBLP:conf/ccs/ZhouWCW14} enables in-process isolation using the, meanwhile deprecated, ARM memory domains (that provide a 4-bit domain ID for isolation).
However, ARM's policy changes need to be performed by the kernel as the policy register is not accessible from user space, incurring additional runtime overhead for policy changes. %

Intel MPK and ARM memory domains only allow for 16 distinct domains, which can be restrictive for certain software architectures that require more in-process domains.
Also, both Intel MPK and ARM memory domains are constrained to the granularity of a page and cannot achieve sub-page granular isolation.
In contrast to protection keys, \tmebox allows the isolation of more in-process domains and provides fine-grained memory isolation on the level of individual cache lines.

Besides MPK, other academic designs~\cite{DBLP:conf/uss/FrassettoJLS18,DBLP:conf/ccs/Xie0ZXLKW022} also leverage page metadata located in the PTE for isolation.
For example, IMIX~\cite{DBLP:conf/uss/FrassettoJLS18} proposes a hardware extension that uses a single bit in the PTE to distinguish between secure and non-secure memory pages.
Access to data located on secure memory pages is only permitted via newly introduced instructions, thereby preventing memory corruption of security-sensitive pages.
CETIS~\cite{DBLP:conf/ccs/Xie0ZXLKW022} proposes the use of Intel CET~\cite{DBLP:conf/isca/ShanbhogueGS19} in order to separate memory resources by placing the isolated memory region on shadow stack pages that can only be accessed with dedicated instructions (\eg \texttt{wrss} instruction). %
These shadow stack pages introduced by Intel CET are identified by an unused combination of the read-write bit and the dirty bit of the PTE.
IMIX and CETIS use their isolation primitive to implement metadata protection, such as code-pointer integrity~(CPI)~\cite{DBLP:conf/osdi/KuznetsovSPCSS14}.
Both countermeasures allow the separation of a process into two distinct domains (\ie secure and non-secure domains) and cannot provide fine-grained isolation (\ie sub-page granular protection).
While this is enough to protect program metadata efficiently, some software systems require numerous in-process sandboxes (as outlined for MPK).

\paragraph{Tag-based Isolation}
Memory tagging technologies associate metadata with data located in memory at a certain granularity, thereby enabling the enforcement of different security policies that restrict access to memory resources~\cite{DBLP:journals/csur/JeroBWSKSO23}.

Tagged architectures such as the ARM memory tagging extension~(MTE)~\cite{DBLP:journals/usenix-login/Serebryany19} and SPARC application data integrity~(ADI)~\cite{DBLP:journals/micro/AingaranJKLLMPR15} hardware features have been efficiently integrated into commercial processors.
ARM MTE is typically used to provide probabilistic memory safety for memory sanitization (\eg MemTagSanatizer) or to help establish runtime security for production software.
However, ARM MTE and SPARC ADI are often criticized for their small tag size of 4-bit (\ie 16 distinct memory tags), resulting in a relatively low detection probability of just 93.75\,\%~\cite{DBLP:journals/corr/abs-1802-09517}.

Furthermore, SFITAG~\cite{DBLP:conf/asiaccs/SeoYCCKP23} combines software instrumentation with ARM MTE to isolate kernel extensions.
However, their design only supports a maximum of 14 isolated memory regions, as two memory tags are reserved.
In contrast, our \tmebox design supports a larger number of concurrent sandboxes (\ie \tmemk is specified for up to 15-bit \keyids that address 32K encryption keys~\cite{intelmktme}).

Also, HAKC~\cite{DBLP:conf/ndss/McKeeGOSPOB22} combines ARM pointer authentication~(PAuth)~\cite{qualcomm} with ARM MTE for kernel compartmentalization.
While HAKC's design expands the possible number of compartments from 4-bit MTE, their security is bound to the cryptographic MAC of the PAuth feature, which depends on the size of the virtual address space.
More specifically, ARM allows the configuration of virtual address sizes between 32 and 52 bits (with bit 55 reserved).
This means that a large virtual address space, as used by servers, can lead to a relatively small MAC, \eg a 52-bit virtual address space results in a MAC size of 11 bits, or even just 3 bits if ARM's top-byte ignore~(TBI) is enabled~\cite{qualcomm}.

\paragraph{Cryptographic Isolation}
Cryptographic primitives, such as encryption and authentication, are used in the context of system security to ensure the confidentiality and integrity of code and data located in memory.
More specifically, these cryptographic primitives can be applied to mitigate the exploitation of memory safety vulnerabilities~\cite{DBLP:conf/ccs/Unterguggenberger23, DBLP:conf/asiaccs/NasahlSWHMM21}.
For instance, cryptographic capability computing~(C$^3$)~\cite{DBLP:conf/micro/LeMayRDDGNGWSGS21} leverages pointer and memory encryption to help prevent memory safety errors.
C$^3$ encrypts the upper address bits of the pointer, creating a cryptographic address~(CA).
Furthermore, C$^3$ encrypts and decrypts accessed data in memory using a keystream generator, with the CA serving as input.
A memory safety error is detected through C$^3$'s pointer decryption by resolving to a garbled address that likely results in a page fault and subsequent program termination.
In contrast to our approach, C$^3$'s proposed hardware modifications directly impact the critical L1 cache latency, whereas \tmebox utilizes Intel \tmemk's DRAM encryption that is available on commodity x86 CPUs.

Intel \tmemk can be used for runtime security to mitigate various attack vectors~\cite{DBLP:conf/eurosp/SchrammelULSGLDM24, DBLP:conf/secrypt/SchrammelSGLDUN23, DBLP:conf/host/NasahlSLGLDSM23, DBLP:journals/csur/ChengOVAGJFB24}.
For example, IntegriTag~\cite{DBLP:conf/eurosp/SchrammelULSGLDM24} uses Intel \tmemk for probabilistic heap memory safety.
IntegriTag uses different security policies (\eg pseudorandom \keyids) that select and implicitly encode a \keyid into the pointer's virtual address and assign the \keyid to the corresponding memory location through aliasing.
This method restricts access to heap objects, as memory accesses are only granted by pointers that incorporate the correct \keyid. %
In this way, \tmemk can be used in the same way as memory tagging, like ARM MTE and SPARC ADI, providing similar security.
Nevertheless, adversaries may still exploit vulnerabilities to leak or guess the \keyid, and to harvest or forge valid pointers, thereby circumventing the security measure.
In contrast, \tmebox enforces the use of the sandbox's \keyid for all memory operations, enabling scalable in-process isolation.
Moreover, EC-CFI~\cite{DBLP:conf/host/NasahlSLGLDSM23} uses Intel \tmemk to thwart fault-induced control-flow hijacking attacks.
EC-CFI encrypts code at the function level, enforcing decryption with different keys for individual functions.
These encryption keys are switched when entering and exiting a function, thereby applying control-flow protection in the presence of fault attacks.
EC-CFI helps to prevent fault-induced control flow hijacking attacks, while our work focuses on software sandboxing.

In addition, Intel trust domain extensions~(TDX)~\cite{DBLP:journals/csur/ChengOVAGJFB24,inteltdxxwhitepaper} enable confidential computing with heavy-weight virtual machine~(VM) isolation.
TDX uses \tmemk for the encryption of VMs and containers~\cite{intelmktmeruntimeencryption}, and the protection against physical attacks~\cite{inteltdxxwhitepaper, DBLP:conf/uss/HaldermanSHCPCFAF08}.
In contrast, \tmebox allows for lightweight and efficient in-process sandboxing without relying on the use of heavy-weight virtualization mechanisms.

\subsection{Possible Extensions}

Software sandboxing applies coarse-grained CFI to restrict control-flow transfers to the sandbox's code region.
\tmebox adheres to conventional SFI techniques in this regard by instrumenting forward-edge and backward-edge control flow transitions.
Other Intel platform-specific hardware features, \ie Intel CET's indirect branch tracking~(IBT) and shadow stack~\cite{DBLP:conf/isca/ShanbhogueGS19}, enable hardware-enforced CFI.
The CET shadow stack feature can be used to ensure the integrity of return addresses.
Applying the CET shadow stack for return address protection within the \tmebox framework would be beneficial, thereby increasing security and optimizing performance.
In addition, CET's IBT limits valid indirect jump targets to the landing pads of function entries, thus hardening against control-flow hijacking attacks.
IBT could also be useful in the context of \tmebox.
However, forward-edge protection still requires instrumentation to enforce that forward-edge transitions remain within the sandbox's code region.
Additionally, FineIBT~\cite{DBLP:conf/raid/GaidisMSMAK23} enforces more fine-grained CFI using IBT.
Here, we see potential synergies between \tmebox and fine-grained CFI to further restrict control-flow transfers.

The \tmebox design can be adopted for the isolation of just-in-time~(JIT) compiled code since in-process isolation is highly demanded in this context, as seen in the V8 sandbox~\cite{v8sandbox,v8sandbox2,v8sandbox3}. %
To achieve this, the JIT compiler must be \tmebox aware to effectively enable our isolation and memory management, \ie the JIT compiler must guarantee that the required instrumentation is performed similarly to our compiler extension.
Moreover, the JIT compiler has to adhere to the Write-XOR-Execute policy~\cite{DBLP:conf/infocom/ZhangNCSCS15,v8sandbox3}, which we assume is enabled for our design.
For instance, the Google Native Client~(NaCl)~\cite{DBLP:conf/sp/YeeSDCMOONF09, DBLP:conf/uss/SehrMBKPSYC10} sandbox has been extended to the JIT compiler of the V8 engine~\cite{DBLP:conf/pldi/AnselMETCSSBY11}. %
We leave it as future work to port \tmebox to JIT compilers.

We use an Intel Xeon Gold 6530 processor as our development and evaluation platform.
While \tmemk is specified for up to 15-bit, it is platform-dependent how many are implemented.
For instance, our processor offers support for 6-bit \keyids.
This means that, excluding the default key (keyID~0), \tmebox supports 63 distinct sandboxes on this CPU.
Nevertheless, we can still support considerably more sandboxes on such processors by using a hybrid solution of \tmebox and SFI.
For instance, in addition to our \keyid instrumentation, we could use additional virtual address bits to isolate coarse-grained SFI regions, where the \tmemk \keyids can be reused for sandboxes in the respective memory region.
This allows us to scale our isolation approach up to thousands of sandboxes on processors that do not implement the full 15-bit \keyids for \tmemk.
Thereby, we increase the number of sandboxes while maintaining the advantages of our design, \ie support for scalable isolation and flexible relocation of data in memory.

\section{Conclusion}\label{sec:tmebox:conclusion}

In this paper, we presented \tmebox, a novel sandboxing technique that provides scalable in-process isolation on commodity x86 CPUs by leveraging Intel's \tmemk memory encryption.
\tmemk is primarily designed to provide page-granular memory encryption for the Intel TDX confidential computing platform, \ie heavy-weight VM isolation.

\tmebox extends the application of \tmemk's integrity enforcement to cryptographically isolate the memory of sandboxes \emph{within} a single process (\ie in-process isolation).
That is, \tmebox uses compiler instrumentation to enforce that the sandboxes use their designated encryption keys for memory interactions, thus detecting unauthorized memory accesses. %
Repurposing \tmemk's runtime encryption for hardware-assisted sandboxing provides unique isolation properties:
\tmebox is scalable, enabling isolation granularities from individual cache lines up to full pages. %
Moreover, our design allows the isolation for up to 32K concurrent sandboxes and offers flexible memory management for the allocator, \ie data relocation in memory.

We prototype our \tmebox framework, consisting of an LLVM compiler toolchain and memory allocator, and propose architecture-specific optimizations, \eg x86 segment-based addressing.
Our performance-optimized prototype demonstrates practical results, showcasing geomean performance overheads of \SPECgs for data and \SPECgscfi for code and data isolation evaluated using the SPEC CPU2017 benchmark suite.

\section*{Acknowledgment}

We thank Andreas Kogler and the anonymous reviewers for their valuable feedback that improved this work.
This project has received funding from the Austrian Research Promotion Agency~(FFG) via the AWARE project (FFG grant number 891092) and the SEIZE project (FFG grant number 888087).
Additional funding was provided by a generous gift from Intel.
Any opinions, findings, and conclusions or recommendations expressed in this paper are those of the authors and do not necessarily reflect the views of the funding parties.

\bibliographystyle{plain}
\bibliography{bibliography.bib}

\end{document}